\documentstyle{article}
\begin{document}
\newcommand{\R}{\bf R}
\newcommand{\K}{\cal K}
\vskip0.5cm
\begin{center}
{\LARGE{\bf \sf Stabilization for equations of one-dimensional}}
\end{center}
\begin{center}
{\LARGE{\bf \sf viscous compressible heat-conducting media}}
\end{center}
\begin{center}
{\LARGE{\bf \sf with nonmonotone equation of state}}
\end{center}
\vskip 0.5cm
\centerline{ {\sf B. Ducomet} $^{a}$
{\it and}    {\sf A.A. Zlotnik} $^{b}$}
\vskip 0.5cm
\centerline{$^{a}$ D\'epartement de Physique Th\'eorique et Appliqu\'ee }
\centerline{CEA/DAM Ile de France}
\centerline{BP 12, F-91680 Bruy\`eres-le-Ch\^atel, {\sc France}}
\centerline{\it{bernard.ducomet@cea.fr }}
\vskip 0.2cm
\centerline{$^{b}$ Department of Mathematical Modelling}
\centerline{Moscow Power Engineering Institute}
\centerline{Krasnokazarmennaja 14, 111250 Moscow, {\sc Russia}}
\centerline{\it{zlotnik@apmsun.mpei.ac.ru}}
\vglue .5truecm
\begin{abstract}
We consider the Navier-Stokes system describing motions of viscous compressible
heat-conducting and ``self-gravitating" media.
We use the state function of the form
$p(\eta,\theta)=p_0(\eta)+p_1(\eta)\theta$ linear with respect to the temperature $\theta$, but we admit
rather general nonmonotone functions $p_0$ and $p_1$
of $\eta$, which allows us to treat various physical models of
nuclear fluids (for which $p$ and $\eta$ are the pressure and specific volume)
or thermoviscoelastic solids.
 For an associated initial-boundary value problem
with ``fixed-free" boundary conditions and possibly large data,
we prove a collection of estimates
independent of time interval for solutions,
including two-sided bounds for $\eta$,
together with its asymptotic behaviour as $t\rightarrow \infty$.
Namely, we establish the stabilization pointwise and in $L^q$  for $\eta$,
in $L^2$ for $\theta$, and in $L^q$ for $v$ (the velocity),
for any $q\in[2,\infty)$.
\end{abstract}

\section{Introduction}

The problem of large-time behaviour of solutions to equations of a 1d-flow of viscous compressible heat-conducting fluids (or gases) with
large data was studied in a lot of papers including \cite{AKM} \cite{Nag} \cite{J} \cite{Q01}. All these papers deal with the case of particular
(polytropic gas) or general pressure law $p(\eta,\theta)$ but always monotone with respect to the variable $\eta$ (here $\eta$ and $\theta$ are
the specific volume and the absolute temperature). It is well known that this monotonicity is not valid in a number of physical situations. In
particular, the case of the two-term pressure
\begin{equation}
p(\eta,\theta)=p_0(\eta)+p_1(\eta)\theta,
\label{i1}
\end{equation}
which is linear in $\theta$ but with complicated nonmonotone
$p_0(\eta)$ is of importance for nuclear fluid models,
see \cite{D01} \cite {D02} and references therein.

The case of the two-term function (\ref{i1}) with other properties of $p_0$
and $p_1$, and nonmonotone $p_1$ is also interesting in
a completely different physical context, namely for thermoviscoelastic solids (shape memory alloys), see \cite{RZ97} \cite{HL98} \cite{SZZ99}
and references therein. In these papers, for models with essentially
simplified forms of the viscosity term and heat flux in the equations, the
stabilization of solutions was studied but
for $\eta$ it was proved only in the case $p_0=0$.

We also mention papers concerning stabilization in nonmonotone barotropic
case (where $p=p(\eta)$) for fluids \cite{K90} \cite{ZB94} \cite{DZ01}
\cite{DZ02} and for viscoelastic solids \cite{AB82} \cite{P87}.

Notice that nonmonotonicity of $p$ complicates in an essential way the problem of stabilization. In particular, the stationary specific volume
becomes nonunique and can be discontinuous.

In this paper, we consider the pressure law (\ref{i1}) with rather general
nonmonotone $p_0$ and $p_1$ and we study both the cases of
nuclear fluids and of thermoviscoelastic solids
(without the aforementioned simplification in the viscosity term and the heat flux).
Moreover a large external force of ``self-gravitation" type is also taken into consideration.
For an initial-boundary value problem with ``fixed-free" boundary conditions
and large initial data,
we prove a collection estimates independent of time interval for solutions, including
two-sided bounds for the specific volume $\eta$. Moreover we establish the pointwise
and $L^q$-stabilization for $\eta$, $L^2$-stabilization for the
temperature $\theta$ and the pressure $p$,
and $L^q$-stabilization for the velocity for any $q\in [2,\infty)$, as time tends to infinity.
In the nuclear fluid case, we also justify the sharpness of the main condition on the ``self-gravitating" force.

\section{Statement of the problem and main results}

 We consider the following system of quasilinear differential equations for 1d-motions of viscous compressible heat-conducting media
\begin{equation}
\left\{  \begin{array}{ll}
\eta_t=v_x, \\[3mm]
v_t=\sigma_x+g, \\[3mm]
e[\eta,\theta]_t=\sigma v_x +\pi_x,
 \end{array}
\right.
\label{e1}
\end{equation}
where $(x,t)\in Q\equiv\Omega\times {\R}^+=(0,M)\times (0,+\infty)$ are
the Lagrangian mass coordinates, with $M$ being the total mass of
the medium.

The unknown quantities $\eta>0$, $v$, and $\theta>0$ are the specific volume,
the velocity, and the absolute temperature.
We also denote by $\rho=\frac{1}{\eta}$ the density, $\sigma=\nu\rho v_x-p[\eta,\theta]$ the stress, $e(\eta,\theta)$ the internal energy, and
$-\pi =-\kappa[\eta,\theta]\rho \theta_x$ the heat flux.

In all the paper, the notation
$\mu[\eta,\theta](x,t)=\mu(\eta(x,t),\theta(x,t))$,
for $\mu=e,p,\kappa,$ etc. is adopted.

In order to fix the state functions $p(\eta,\theta)$ and $e(\eta,\theta)$,
we define the Helmholtz free energy
$$\Psi(\eta,\theta)=-c_V \theta \log \theta-P_0(\eta)-P_1(\eta)\, \theta,$$
where $c_V=const>0$. Then thermodynamics tells us that
\begin{equation}
p(\eta,\theta)=-\Psi_{\eta}(\eta,\theta)=p_0(\eta)+ p_1(\eta)\, \theta ,
\label{e5}
\end{equation}
with $p_0=P_0'$ and $p_1=P_1'$, as well as
\begin{equation}
e(\eta,\theta)=\Psi(\eta,\theta)-\theta \Psi_{\theta}(\eta,\theta)=-P_0(\eta)+c_V \theta ,
\label{e6}
\end{equation}
where $\Psi_{\eta}=\frac{\partial\Psi}{\partial\eta}$ and $\Psi_{\theta}=\frac{\partial\Psi}{\partial\theta}$.

First, we consider the more difficult case of the nuclear fluid.
We suppose that the functions $p_0,p_1 \in C^1({\R}^+)$ are such that
\footnote{Note that $C^1({\R}^+)$ stands for the space of
continuously differentiable functions on ${\R}^+$, but not necessarily bounded.
The spaces $C^1({\R}^+ \times {\R}^+)$,  $C({\R}^+)$, $C({\R})$, etc.  used
below are understood similarly.}
\begin{equation}
{\displaystyle \lim_{\eta\rightarrow 0^+}p_0(\eta)=+\infty,\ \ \lim_{\eta\rightarrow \infty}p_0(\eta)=0,}
\label{p0}
\end{equation}
\begin{equation}
{\displaystyle
p_1(\eta)\geq 0,\ \ \eta\, p_1(\eta)=O(1)\ \mbox{as}\ \eta\rightarrow \infty.}
\label{p1}
\end{equation}
Suppose also that the viscosity and heat conductivity coefficients are such that
$\nu=const>0$ and $\kappa\in C^1({\R}^+\times {\R}^+)$,
with $0<\underline{\kappa}\leq \kappa(\eta,\theta)\leq \overline{\kappa}$, where
$\underline{\kappa}$ and $\overline{\kappa}$
are given constants.

Let the so-called ``self-gravitation force" $g\in L^1(\Omega)$. In fact, this name does not correspond exactly to the physical
situation, as, at least in the nuclear fluid case, the corresponding ``physical" force is the Coulomb force between charged particles, which
contrary to the Newton gravitational force, is attractive.
Although the distinction Coulomb-Newton is of utmost importance in multidimensional
problems, it is harmless in the 1d-context.

Let us supplement equations (\ref{e1}) with the following boundary and initial conditions
\begin{equation}
\left.v\right|_{x=0}=0,\ \left. \sigma\right|_{x=M}=-p_{\Gamma},\ \left.\theta\right|_{x=0}=\theta_{\Gamma},\ \left.\pi\right|_{x=M}=0,
\label{e7}
\end{equation}
\begin{equation}
\left. \eta\right|_{t=0}=\eta^0(x),\ \ \left.v\right|_{t=0}=v^0(x), \ \ \left.\theta\right|_{t=0}=\theta^0(x),
\label{e8}
\end{equation}
with an outer pressure $p_{\Gamma}=const$ and a given temperature $\theta_{\Gamma}=const>0$.

Throughout the paper, we use the classical Lebesgue spaces $L^q(G)$ together with their anisotropic version $L^{q,r}(Q)$,
for $q,r\in [1,+\infty]$, and we denote
the associated norm by $\|\cdot\|_{L^{q,r}(Q)}=\left\|\ \|\cdot\|_{L^q(\Omega)}\ \right\|_{L^r({\R}^+)}$.
In Section 2, we also use the abbreviate notation $\|\cdot\|_G$ for $\|\cdot\|_{L^2(G)}$.

Let also $V_2(Q)$ be the standard space of functions $w$ having finite (parabolic) energy
$\|w\|_{V_2(Q)}=\|w\|_{L^{2,\infty}(Q)}+\|w_x\|_{L^2(Q)}$.
We denote by $H^1(\Omega)$ (resp. $H^{2,1}(Q_T)$) the standard Sobolev space
equipped with the norm
$\|\phi\|_{H^1(\Omega)}=\|\phi\|_{L^2(\Omega)}+\|\phi_x\|_{L^2(\Omega)}$ (resp.
$\|w\|_{H^{2,1}(Q_T)}=\|w\|_{L^{2,\infty}(Q_T)}+\|w_x\|_{V_2(Q_T)}+\|w_t\|_{L^2(Q_T)}$).
Hereafter $Q_T=\Omega\times (0,T)$.

In Section 2, we shall also exploit the integration operators $I^*\phi(x)=\int_x^M \phi(\xi)\, d\xi$, for $\phi\in L^1(\Omega)$, and
$I_0a(t)=\int_0^t a(\tau)\, d\tau,$ for $a\in L^1(0,T).$
\vskip0.5cm
Suppose that the initial data are such that $\eta^0\in L^{\infty}(\Omega)$ with $\mbox{ess inf}_{\Omega}\eta^0>0$, $v^0\in L^4(\Omega)$,
$\theta^0\in L^2(\Omega)$, $\log \theta^0\in L^1(\Omega)$ with $\theta^0>0$.

Though it is possible to establish our main results for weak solutions \cite{AZ89}, to simplify the presentation, we limit ourselves to the case of
so-called regular weak solutions \cite{AKM} such that
$\eta\in L^{\infty}(Q_T)$, $\eta_x,\eta_t \in L^{2,\infty}(Q_T)$,
$\mbox{min}_{\overline{Q}_T}\eta>0$, and
$v, \theta \in  H^{2,1}(Q_T)$, $\theta>0$ in $Q_T$, for any $T>0$.
We consider the problem of existence of the latter solutions in Appendix.

Now we summarize our main results concerning problem (\ref{e1}), (\ref{e7}),
(\ref{e8}), under conditions (\ref{p0}),(\ref{p1}).

Let us define the function
$$p_S(x):=p_{\Gamma}-\int_x^M g(\xi)\, d\xi,\ \
\mbox{for}\ x\in\overline{\Omega},$$
which plays the role of a stationary pressure,
and set $\underline{p}_{\, S}:=\mbox{min}_{\, \overline{\Omega}}\ p_S$ and
$\overline{p}_S:=\mbox{max}_{\, \overline{\Omega}}\ p_S$. Obviously
$\underline{p}_{\, S}\leq p_{\Gamma}\leq \overline{p}_S$.

Let $N>1$ be an arbitrarily large parameter and $K_i=K_i(N)$ and
$K^{(i)}=K^{(i)}(N)$, $i=0,1,2,...$, be positive nondecreasing functions of $N$,
which can also depend on $M,\nu,\underline{\kappa},\overline{\kappa}, \mbox{etc}$, but neither on the initial data nor on $g$.

\newtheorem{th1}{Theorem}
 \begin{th1}
1. Suppose that the initial data, $p_{\Gamma},$ and $g$ are such that
\begin{equation}
N^{-1}\leq \eta^0\leq N,\ \ \|v^0\|_{L^4(\Omega)}+\|\log \theta^0\|_{L^1(\Omega)}+\|\theta^0\|_{L^2(\Omega)}\leq N,
\label{t1}
\end{equation}
\begin{equation}
\|g\|_{L^1(\Omega)}\leq N,\ \ N^{-1}\leq \underline{p}_{\, S}.
\label{t2}
\end{equation}
Then the following estimates in $Q$ together with $L^2(\Omega)$-stabilization property hold
\begin{equation}
0<K_1^{-1}=\underline{\eta}\leq \eta(x,t) \leq \overline{\eta}=K_2\ \
\mbox{in}\ \overline{Q},
\label{t5}
\end{equation}
\begin{equation}
\|v\|_{V_2(Q)}+\|v^2\|_{V_2(Q)}+\|\log \theta\|_{L^{1,\infty}(Q)}+\|(\log \theta)_x \|_{L^2(Q)}+\|\theta-\theta_{\Gamma}\|_{V_2(Q)}\leq K_3,
\label{t6}
\end{equation}

$$\|p[\eta,\theta]-p_S\|_{L^2(Q)}\leq K_4,$$

$$\|v^2(\cdot,t)\|_{L^2(\Omega)}+\|\theta(\cdot,t)-\theta_{\Gamma}\|_{L^2(\Omega)}
+\|p[\eta,\theta](\cdot,t)-p_S(\cdot)\|_{L^2(\Omega)}\rightarrow 0
\ \ \mbox{as}\ t\rightarrow \infty.$$

2. Suppose that $p(\eta,\theta)$ satisfies the following
additional condition:
\begin{equation}
\mbox{For any}\ c\in[\underline{p}_{\, S},\overline{p}_S],\ \mbox{there exists no interval}\ (\eta_1,\eta_2)\ \mbox{such that}\
p(\eta,\theta_{\Gamma})\equiv c\ \  \mbox{on}\ (\eta_1,\eta_2).
\label{t3}
\end{equation}
Then the following pointwise stabilization property holds for $\eta$: there exists a function $\eta_S\in L^{\infty}(\Omega)$ satisfying
\begin{equation}
p(\eta_S(x),\theta_{\Gamma})=p_S(x)\ \  \mbox{and}\ \ \underline{\eta}
\leq \eta_S(x)\leq \overline{\eta}\ \mbox{on}\ \overline{\Omega},
\label{t4}
\end{equation}
such that
\begin{equation}
\eta(x,t) \rightarrow \eta_S(x)\ \ \mbox{as}\ t\rightarrow \infty,
\ \ \mbox{for all}\ x\in\overline{\Omega}.
\label{t7}
\end{equation}
and consequently $\|\eta(\cdot,t)-\eta_S(\cdot)\|_{L^q(\Omega)}\rightarrow 0$ as $t\rightarrow \infty$, for any $q\in[1,\infty)$.

3. Suppose that, additionally to the hypotheses of Claim 1,
$\|v^0\|_{L^q(\Omega)}\leq N$, for some $q\in(4,\infty)$.
Then the following estimate  in $Q$ together with $L^q(\Omega)$-stabilization property hold
$$\|v\|_{L^{q,\infty}(Q)}+\|v\|_{L^{\infty,q}(Q)}\leq K_5q,$$
$$\|v(\cdot,t)\|_{L^q(\Omega)}\rightarrow 0 \ \ \mbox{as}\ t\rightarrow \infty,$$
where $K_5$ does not depend on $q.$
 \label{th1}
\end{th1}

{\bf Remarks:}

1. An elementary but important consequence of Claim 2 is that $V(t):=\int_{\Omega}\eta(x,t)\, dx\rightarrow V_S>0$ as
$t \rightarrow \infty$, where $V(t)$ is the volume of the fluid (or in other words, the Eulerian position of the free boundary).

2. For nonmonotone $p(\eta,\theta_{\Gamma})$, if there exist two points $0<\eta^{(1)}<\eta^{(2)}$ such that
$$\underline{p}_{\, S}<p^{(1)}:=p(\eta^{(1)},\theta_{\Gamma})<p^{(2)}:=p(\eta^{(2)},\theta_{\Gamma})
<\overline{p}_S,$$
and such that, moreover

$$\left\{  \begin{array}{ll}
p^{(1)}\leq p(\eta,\theta_{\Gamma}),\ \ \mbox{for}\ 0<\eta\leq \eta^{(1)},\\[3mm]
p^{(1)}\leq p(\eta,\theta_{\Gamma})\leq p^{(2)},\ \ \mbox{for}\ \eta^{(1)}<\eta<\eta^{(2)},\\[3mm]
p(\eta,\theta_{\Gamma})\leq p^{(2)},\ \ \mbox{for}\ \eta^{(2)}\leq \eta,
 \end{array}
\right.$$
then, necessarily $\eta_S\notin C(\overline{\Omega})$. Moreover, consequently, the convergence in (\ref{t7}) cannot be uniform in $x$.
In fact, even for $g\equiv 0$, if the equation $p(\eta,\theta_{\Gamma})=p_{\Gamma}$ has
more than one solution, then $\eta_S$ can be discontinuous in $\overline{\Omega}$. Namely, if this equation has exactly $k$ solutions
$\eta^{(1)}<...<\eta^{(k)}$, then the function $\eta_S$ can be written as
$$\eta_S=\sum_{j=1}^k \chi(E_j)\, \eta^{(j)},$$
where $E_j,\, 1\leq j \leq k,$ are any measurable nonintersecting subsets
of $\overline{\Omega}$ (some of them may be empty) such that
${\displaystyle \bigcup_{j=1}^k E_j=\overline{\Omega}}$, and $\chi(E_j)$ are their characteristic functions. Unfortunately, we cannot assert
more about $\eta_S$.

Let us justify that the second condition (\ref{t2}) is essential in Theorem 1. Set $m(\theta_{\Gamma}):=\inf_{\eta>0}p(\eta,\theta_{\Gamma})$.
Obviously $m(\theta_{\Gamma})\leq0$, and if $p_0\geq -p_1\theta_{\Gamma}$, then $m(\theta_{\Gamma})=0$.

\newtheorem{prop1}{Proposition}
 \begin{prop1}
Let the hypotheses of theorem 1, Claim 1, be valid, but suppose that
$\underline{p}_{\, S}< m(\theta_{\Gamma})$, instead of
$N^{-1}\leq \underline{p}_{\, S}$.
Then

\begin{equation}
\limsup_{t\rightarrow \infty} V(t)=\infty.
\label{pr11}
\end{equation}

\label{prop1}
\end{prop1}

This property means that the upper bound for $\eta$ in (\ref{t5}) is violated and physically, that the fluid can asymptotically expand in
the whole halfspace.
\vskip0.2cm
Let us also consider the borderline case $\underline{p}_{\, S}= m(\theta_{\Gamma})$.

\newtheorem{prop2}[prop1]{Proposition}
 \begin{prop2}

Let the hypotheses of theorem 1, Claim 1, be valid and
$p(\eta,\theta_{\Gamma})> m(\theta_{\Gamma})=0$, but $p_S(0)=0 $ instead of
$N^{-1}\leq p_S$. Then at least one of the following properties holds:

\begin{equation}
\limsup_{t\rightarrow \infty} \left|\int_{\Omega} v(x,t)\, dx\right|=\infty,
\label{pr21}
\end{equation}

\begin{equation}
\lim_{t\rightarrow \infty} \eta(0,t)=\infty .
\label{pr22}
\end{equation}

If in addition $\int_1^{\infty}p(\eta,\theta_{\Gamma})\, d\eta<\infty$ and
$\underline{p}_{\, S}=p_S(0)=0$, then
$\|v\|_{L^{2,\infty}(Q)}\leq K_3$, but property (\ref{pr22}) holds.
\label{prop2}
\end{prop2}

Properties (\ref{pr21}) and (\ref{pr22}) mean that estimate (\ref{t6}) and the upper bound for $\eta$ in (\ref{t5}) are violated respectively.

Note that propositions \ref{prop1} and \ref{prop2} go back to results of
\cite{ZB94} where the barotropic case was studied.
\vskip0.5cm
Finally, we consider the case of thermoviscoelastic solids.
Let $\underline{p}_{\, S}\leq \overline{p}_S$ be fixed. Suppose that, instead of (\ref{p0}) and (\ref{p1}), the following conditions hold
\begin{equation}
{\displaystyle \overline{p}_S\leq p_0(\eta)\ \mbox{and}\ 0\leq p_1(\eta)\ \mbox{for}\ 0<\eta\leq \check{\eta},}
\label{p2}
\end{equation}
\begin{equation}
{\displaystyle p_0(\eta)\leq \underline{p}_{\, S}\ \mbox{and}\ p_1(\eta)\leq 0 \ \mbox{for}\ 0<\hat{\eta}\leq \eta,}
\label{p3}
\end{equation}
for some $0<\check{\eta}\leq\hat{\eta}<\infty$.
The conditions of such kind are of standard type for the thermoviscoelastic case.

\newtheorem{th2}[th1]{Theorem}
 \begin{th2}

 All the Claims 1-3 of theorem \ref{th1} remain valid under conditions (\ref{p2}) and (\ref{p3}), and without the condition
$N^{-1}\leq \underline{p}_{\, S}$.
\label{th2}
\end{th2}
{\bf Remark:}

 We could consider the viscosity coefficient
$\nu=\nu(\eta)\geq \nu_0>0$, $\nu\in C^1({\R}^+)$ as well as body force and boundary data in the form
 $g(x,t)=g_S(x)+\Delta g(x,t)$,
$p_{\Gamma}(t)=p_{\Gamma,S}+\Delta p_{\Gamma}(t)$, and
$\theta_{\Gamma}(t)=\theta_{\Gamma,S}+\Delta \theta_{\Gamma}(t)$, with
perturbations $\Delta g, \Delta p_{\Gamma},$ and $\Delta \theta_{\Gamma}$
tending to zero as $t\rightarrow \infty$ in some weak sense (compare with the barotropic case \cite{ZB94} \cite{DZ02}).
To simplify the presentation of the results and their proof, we do not realize this possibility in the paper.

\section{Proof of the results }

We begin with the proof of theorem \ref{th1} which follows from a lengthy series of lemmas, providing necessary
a priori estimates and stabilization properties: Claims 1, 2, and 3 will be proved respectively in lemmas 1-9, lemmas 10 and 11, and lemmas 12
and 13.

Then we proceed with the proofs of propositions \ref{prop1} and \ref{prop2} and theorem \ref{th2}.

\subsection{A priori estimates and proof of theorem 1}

\newtheorem{lem1}{Lemma}
 \begin{lem1}
The following energy estimates hold
\begin{equation}
\|\eta\|_{L^{1,\infty}(Q)}+\|v\|_{L^{2,\infty}(Q)}+\|\theta\|_{L^{1,\infty}(Q)}+\|\log \theta\|_{L^{1,\infty}(Q)}\leq K^{(1)},
\label{l11}
\end{equation}
\begin{equation}
\|\sqrt{\frac{\rho}{\theta}} v_x\|_Q+\|\frac{\sqrt{\rho}}{\theta} \theta_x\|_Q \leq K^{(2)}.
\label{l12}
\end{equation}
\label{lem1}
\end{lem1}
{\bf Proof:} Equations (\ref{e1}) and (\ref{e5}), (\ref{e6}) imply the equations
\begin{equation}
\left(\frac{1}{2}v^2+e[\eta,\theta]\right)_t=(\sigma v+\pi)_x+gv,
\label{l13}
\end{equation}
\begin{equation}
c_V\theta_t=\pi_x+\left(\nu \rho v_x-p_1[\eta]\theta\right) v_x.
\label{l13a}
\end{equation}
Hereafter we use the notation $\mu[\eta](x,t)=\mu(\eta(x,t))$,
for $\mu=p_i,P_i,\  i=0,1$, etc.

By multiplying the second equation by $\frac{1}{\theta}$, we get
$$\left(c_V \log \theta +P_1[\eta]\right)_t=\frac{1}{\theta}\pi_x+\nu \frac{\rho}{\theta} v_x^2.$$
By multiplying this relation by $\theta_{\Gamma}$, and subtracting from equation (\ref{l13}), we obtain
$$\left(\frac{1}{2}v^2+e[\eta,\theta]-c_V \theta_{\Gamma}\log \frac{\theta}{\theta_{\Gamma}}-\theta_{\Gamma}P_1[\eta]+p_{\Gamma}\eta \right)_t
+\theta_{\Gamma}\nu \frac{\rho}{\theta} v_x^2
=\left((\sigma +p_{\Gamma})v\right)_x+\left(1-\frac{\theta_{\Gamma}}{\theta}\right)\pi_x+gv.$$
Setting $P(\eta, \theta):=P_0(\eta)+P_1(\eta)\, \theta,$
integrating this equality over $\Omega$, and using the formula
$$\int_{\Omega} gv\, dx=\int_{\Omega} \left( I^*g \right) v_x\, dx=\frac{d}{dt}\int_{\Omega} \left( I^*g \right) \eta\, dx,$$
we finally get
$$\frac{d}{dt}\int_{\Omega}
\left[\frac{1}{2}v^2+c_V \theta_{\Gamma} \left(\frac{\theta}{\theta_{\Gamma}}-\log \frac{\theta}{\theta_{\Gamma}}\right)
+p_S\eta-P[\eta, \theta_{\Gamma}]+C\right]\, dx$$
\begin{equation}
+\theta_{\Gamma}\int_{\Omega}\left( \nu \frac{\rho}{\theta}v_x^2+ \kappa[\eta,\theta]\frac{\rho}{\theta^2}\theta_x^2 \right)\, dx=0,
\label{l14}
\end{equation}
for any constant $C$.

Conditions (\ref{p0}) and (\ref{p1}) imply the property
$$P(\eta,\theta_{\Gamma})\leq \varepsilon \eta+C_{\varepsilon}\ \ \mbox{on}\ {\R}^+,\ \forall \varepsilon>0.$$
By integrating (\ref{l14}) over $(0,T)$ for any $T>0$, applying conditions (\ref{t1}) and (\ref{t2}) and choosing
$\varepsilon:=\frac{1}{2}\underline{p}_{\, S}$,
we obtain estimates (\ref{l11}) and (\ref{l12}). Here, the elementary inequality $\frac{1}{2}\alpha \leq \alpha -\log \alpha+\log 2-1$ is taken
into account.$\ \ \ \Box$
\vskip0.2cm
The following auxiliary result on ordinary differential inequalities is useful to prove lower and upper bounds for the specific volume $\eta$
in various situations.

\newtheorem{lem2}[lem1]{Lemma}
 \begin{lem2}
Let $N_0\geq 0$, $N_1\geq 0$, and $\varepsilon_0>0$ be three parameters.

Let $f\in C({\R})$ and $y,b\in W^{1,1}(0,T)$, for any $T>0$.
The following claims are valid:

1. if $$\frac{dy}{dt}\geq f(y)+\frac{db}{dt}\ \mbox{on}\ {\R}^+,$$
where $f(-\infty )=+\infty$ and $b(t)-b(\tau )\geq -N_0-N_1(t-\tau )$, for any $0\leq \tau \leq t$, then the uniform lower bound holds:
$$\min \{ y(0),\check{z} \}
-N_0 \leq y(t)
\ \mbox{on}
\ \overline{ {\R} }^+,$$
where the number $\check{z}=\check{z}(N_1)$ is such that $f(z)\geq N_1$, for $z\leq \check{z}$;

2. if $$\frac{dy}{dt}\leq f(y)+\frac{db}{dt}\ \mbox{on}\ {\R}^+,$$
where $\limsup_{z\rightarrow +\infty} f(z)\leq 0$,
and $b(t)-b(\tau)\leq N_0-\varepsilon_0(t-\tau)$, for any $0\leq \tau \leq t$,
then the uniform upper bound holds:
$$y(t)\leq \max \{ y(0),\hat{z} \}+N_0\ \mbox{on}\ \overline{ {\R} }^+,$$
where the number $\hat{z}=\hat{z}(\varepsilon_0)$ is such that $f(z)\leq \varepsilon_0$, for $z\geq \hat{z}$.
\label{lem2}
\end{lem2}
{\bf Remark:}
\vskip0.2cm
In lemma \ref{lem2}, one can drop the conditions $f(-\infty)=+\infty$ and
$\limsup_{z\rightarrow +\infty} f(z)\leq 0,$
take $f\in C({\R}\times {\R}^+)$ and replace $f(y)$ by $f(y,t)$. Then Claim 1 remains valid if, for a fixed $N_1$, there
exists $\check{z}$ such that $f(z,t)\geq N_1$, for $z\leq \check{z}$ and $t\geq 0$. Similarly, Claim 2 remains valid if, for a fixed
$\varepsilon_0\geq 0$, there exists $\hat{z}$ such that $f(z,t)\leq \varepsilon_0$, for $z\geq \hat{z}$ and $t\geq 0$.
\vskip0.5cm
Lemma \ref{lem2} is borrowed from \cite{Z00}, where in both claims, differential equalities are used, but one checks easily that the proof
remains valid for inequalities; the similar conclusion is valid concerning the above remark. The statements of the type specified in this remark are well
known in viscoelastic and thermoviscoelastic contexts.

\newtheorem{lem3}[lem1]{Lemma}
 \begin{lem3}
For $\eta$, the uniform lower bound holds
$$0<\underline{\eta}=\left(K^{(3)}\right)^{-1}\leq \eta(x,t) \
\mbox{in}\ \overline{Q}.$$

 \label{lem3}
\end{lem3}
{\bf Proof:} The action of the operator $I^*$ on the second equation (\ref{e1})
gives the equation
\begin{equation}
I^*v_t=-\nu \rho v_x+p[\eta,\theta]-p_S,
\label{l30}
\end{equation}
which together with the relation $\rho v_x=(\log \eta)_t$ lead to the
another important equation
\begin{equation}
(\nu \log \eta)_t=p[\eta,\theta]-p_S-I^*v_t.
\label{l31}
\end{equation}
By putting $y:=\nu \log \eta$, exploiting the property $p_1[\eta]\theta\geq 0$, and fixing any
$x\in \overline{\Omega}$, we get
$$\frac{dy}{dt}\geq p_0\left(\exp(\frac{y}{\nu})\right)-\overline{p}_S-\frac{d}{dt} I^*v.$$
The function $f(z):=p_0\left(\exp(\frac{z}{\nu})\right)-\overline{p}_S$
satisfies the property $f(-\infty)=+\infty$ (see (\ref{p0})).
Moreover, due to the energy estimate (\ref{l11})
\begin{equation}
\left| \left. I^*v\right|_{\tau}^t\right| \leq 2\ \sup_{\overline{Q}}\left|I^*v\right|\leq 2M^{1/2} \|v\|_{L^{2,\infty}(Q)}\leq K_0.
\label{l32}
\end{equation}
Now Claim 1 in lemma \ref{lem2} (with $N_1=0$) implies the estimate
$$\min \{\nu \log \eta^0(x), \nu \log \check{\eta}\}-K_0\leq y(x,t),$$
with a number $\check{\eta}$ such that $p_0(\eta)-\overline{p}_S\geq 0$, for any $0<\eta \leq \check{\eta}$. Then:
$$\underline{\eta}:=\min \{ N^{-1},\check{\eta}\}\, \exp(-\frac{K_0}{\nu})
\leq \eta(x,t)\ \mbox{in}\ \overline{Q}.\ \ \ \Box$$

The next auxiliary result on ordinary integral inequality is useful to deduce a uniform upper bound for $\eta$.

\newtheorem{lem4}[lem1]{Lemma}
 \begin{lem4}
Let $b$ be a nondecreasing function on $[0,T]$ with $b(0)\geq 0$,
and let $a\in L^1(0,T)$ be a nonnegative function.
If $z\in L^{\infty}(0,T)$, $z\geq 0$ satisfies
$$z(t)\leq b(t)+\int_0^t a(\tau) z(\tau)\, d\tau \ \mbox{on}\ (0,T),$$
then the upper bound holds:
$$z(t)\leq b(t)\exp\left(\int_0^ta(\tau) \, d\tau\right) \leq b(t)\ \exp\left( \|a\|_{L^1(0,T)}\right) \ \mbox{on}\ (0,T).$$

 \label{lem4}
\end{lem4}

The result follows immediately from the integral Gronwall's lemma
(for example see \cite{AKM}) if one takes into account that
$$z(s)\leq b(t)+\int_0^s a(\tau) z(\tau)\, d\tau, \ \mbox{for}\ 0<s\leq t<T.$$

\newtheorem{lem5}[lem1]{Lemma}
 \begin{lem5}
For $\eta$, the uniform upper bound holds
$$\eta(x,t)\le\overline{\eta}=K^{(4)} \ \mbox{in}\ \overline{Q}.$$
 \label{lem5}
\end{lem5}
{\bf Proof:} Let us rewrite the first equation (\ref{e1}) as follows
  $$\eta_t=\frac{1}{\nu} (\sigma+\delta)\ \eta +\frac{1}{\nu}\eta  \left(p[\eta,\theta]-\delta\right),$$
  where $\delta$ is a parameter.
We consider this equation as an ordinary differential equation
with respect to $\eta$ and obtain the formula
\begin{equation}
\eta=\exp \left( \frac{1}{\nu}\ I_0(\sigma+\delta) \right) \left\{
\eta^0+\frac{1}{\nu}I_0 \left[ \exp\left(-\frac{1}{\nu}\
I_0(\sigma+\delta)\right)  \eta\left(p[\eta,\theta]-\delta\right)
\right]\right\}. \label{l51}
\end{equation}
By applying the operator $I_0$ to equation (\ref{l30}), we find
$$I_0\sigma=-p_St-I^*(v-v^0).$$
So by choosing $\delta:=\frac{1}{2} \underline{p}_{\,S}$ and using estimate (\ref{l32}), we get
$$\frac{1}{\nu}\ \left. I_0(\sigma+\delta) \right|_{\tau}^t
=-\frac{1}{\nu}\ (p_S-\delta)(t-\tau)-\frac{1}{\nu}\left.
I^*v \right|_{\tau}^t
\leq -\alpha(t-\tau)+K_1\ \ \mbox{on}\ \overline{\Omega},
\ \ \mbox{for all}\ 0\leq \tau \leq t
$$
with $\alpha:=\frac{1}{2\nu}\underline{p}_{\,S}>0$. Conditions (\ref{p0}) and (\ref{p1}) on $p_0$ and $p_1$ together with the lower bound
$\underline{\eta}\leq \eta$ give
$$\eta\left(p[\eta,\theta]-\delta\right) \leq \eta\, \max\{p_0[\eta]-\delta,0\}+\eta p_1[\eta]\theta
\leq K_2+K_3\theta.$$
Therefore formula (\ref{l51}) implies the estimate
\begin{equation}
\hat{\eta}(t):=\|\eta(\cdot,t)\|_{L^{\infty}(\Omega)}\leq K_4 e^{-\alpha t}\left[1
+\int_0^t e^{\alpha \tau}\left(1+\|\theta(\cdot,\tau)\|_{L^{\infty}(\Omega)}\right)\, d\tau \right].
\label{l52}
\end{equation}
Set $a:=\left\|\frac{\sqrt{\rho}}{\theta}\theta_x\right\|_{\Omega}^2$.
It is well known \cite{AKM} \cite{AZ92} that the inequalities
$$\|\theta\|_{L^{\infty}(\Omega)}
\leq \frac{1}{M} \|\theta\|_{L^1(\Omega)}+\|\theta_x\|_{L^1(\Omega)}
\leq \frac{1}{M} \|\theta\|_{L^1(\Omega)}+\left(a\|\theta\|_{L^1(\Omega)} \|\theta\|_{L^{\infty}(\Omega)}\hat{\eta}\right)^{1/2}$$
$$\leq \varepsilon \|\theta\|_{L^{\infty}(\Omega)}+\frac{1}{M} \|\theta\|_{L^1(\Omega)}+\frac{1}{4\varepsilon} a \|\theta\|_{L^1(\Omega)}\hat{\eta}
,\ \forall \varepsilon>0$$
together with the estimate $\|\theta\|_{L^{1,\infty}(Q)}\leq K^{(1)}$ imply
$$\|\theta\|_{L^{\infty}(\Omega)}\leq K_5\left( 1+a\hat{\eta}\right).$$
So by using estimate (\ref{l52}), the function $z(t):= e^{\alpha t} \hat{\eta}(t)$ satisfies
$$z(t)\leq K_6\left( e^{\alpha t} +\int_0^t a(\tau) z(\tau)\, d\tau \right)\ \ \mbox{on}\ {\R}^+.$$
As $\|a\|_{L^1({\R}^+)}\leq (K^{(2)})^2$ according to lemma \ref{lem1}, by using lemma \ref{lem4}
$$z(t)\leq K_6\, \exp \left( \alpha t+K_6(K^{(2)})^2\right)=K_7 e^{\alpha t}\ \ \mbox{on}\ {\R}^+.$$
This means that
$\eta\leq \hat{\eta}\leq \overline{\eta}:=K_7 \ \ \mbox{in}\ \overline{Q}.\ \ \ \ \Box$

 \newtheorem{cor1}{Corollary}
 \begin{cor1}
For $v$, the following estimate holds
$$\frac{1}{\sqrt{M}} \|v\|_Q\leq \|v\|_{L^{\infty,2}(Q)}\leq (K^{(1)})^{1/2}\left\|\frac{v_x}{\sqrt{\theta}}\right\|_Q\leq K^{(5)}.$$
 \label{cor1}
\end{cor1}
{\bf Proof:} In fact, by using lemma \ref{lem1}, we have

\begin{equation}
   \|v\|_{C(\overline{\Omega})}
\leq \|v_x\|_{L^1(\Omega)}
\leq \|\theta\|_{L^1(\Omega)}^{1/2}
\left\|\frac{v_x}{\sqrt{\theta}}\right\|_\Omega
\leq (K^{(1)})^{1/2}\left\|\frac{v_x}{\sqrt{\theta}}\right\|_\Omega,
\label{c11}
\end{equation}
and
$$\left\|\frac{v_x}{\sqrt{\theta}}\right\|_Q
\leq \overline{\eta}^{1/2}\left\|\sqrt{\frac{\rho}{\theta}}v_x \right\|_Q
\leq \overline{\eta}^{1/2}K^{(2)}.\ \ \ \ \Box$$

Note that similarly
$\| (\log \theta)_x\|_Q\leq \overline{\eta}^{1/2}K^{(2)}.$

\vskip0.2cm
The following auxiliary result on ordinary differential inequalities will be exploited when proving $V_2(Q)$-estimates and
$L^2(\Omega)$-stabilization for $v^2$ and $\theta-\theta_{\Gamma}$.

\newtheorem{lem6}[lem1]{Lemma}
 \begin{lem6}
Let $a_0=const>0$ and $a,h\in L^1({\R}^+)$. If a function $y\geq 0$ on ${\R}^+$ satisfies $y\in W^{1,1}(0,T)$ for any $T>0$ and
\begin{equation}
\frac{dy}{dt}+(a_0+a)y\leq h\ \ \mbox{on}\ {\R}^+,
\label{l61}
\end{equation}
then the following upper bound together with stabilization property hold:
\begin{equation}
y(t)\leq \left(y(0)+\|h\|_{L^1({\R}^+)}\right)\ \
exp \left(\|a\|_{L^1({\R}^+)}\right) \ \ \mbox{on}\ {\R}^+,
\label{l62}
\end{equation}
$$y(t)\rightarrow 0\ \mbox{as}\ t\rightarrow \infty. $$
 \label{lem6}
\end{lem6}

It is easy to derive this simple known result by multiplying (\ref{l61}) by $\exp I_0(a_0+a)$ and integrating the result; of course estimate
(\ref{l62}) holds also for $a_0=0$. Note that more general result can be found in \cite{SZ01}, lemma 2.1.

\newtheorem{lem7}[lem1]{Lemma}
 \begin{lem7}
 For $v^2$ and $\theta-\theta_{\Gamma}$, the following estimate with the stabilization property hold
 $$\|v^2\|_{V_2(Q)}+\|\theta-\theta_{\Gamma}\|_{V_2(Q)}\leq K^{(6)},$$
 \begin{equation}
\|v^2(\cdot,t)\|_{\Omega}+\|\theta(\cdot,t)-\theta_{\Gamma}\|_{\Omega}
\rightarrow 0\ \ \mbox{as}\ t\rightarrow \infty.
\label{l70}
\end{equation}
 \label{lem7}
\end{lem7}
{\bf Proof:} By rewriting equation (\ref{l13}) as follows
  $$\left( \frac{1}{2} v^2+c_V(\theta-\theta_{\Gamma})\right)_t
=(\sigma v+\pi)_x+p_0[\eta]v_x+gv$$
and taking $L^2(\Omega)$-inner product with $\frac{1}{2} v^2+c_V(\theta-\theta_{\Gamma})$, we obtain
  $$\frac{1}{2} \frac{d}{dt} \int_{\Omega} \left( \frac{1}{2} v^2+c_V(\theta-\theta_{\Gamma})\right)^2\, dx
  +\int_{\Omega}\left[ (\nu \rho v_x-p[\eta,\theta])v+\kappa[\eta,\theta]\rho \theta_x\right]
\left( vv_x+c_V\theta_x\right)\, dx$$
  \begin{equation}
=\int_{\Omega} \left( p_0[\eta]v_x+gv\right) \left( \frac{1}{2} v^2+c_V (\theta-\theta_{\Gamma})\right)\, dx
-p_{\Gamma}\left.
\left( v \left(\frac{1}{2} v^2+c_V(\theta-\theta_{\Gamma})\right)\right)\right|_{x=M}.
\label{l71}
\end{equation}
We also take $L^2(\Omega)$-inner product of the second equation (\ref{e1})
with $v^3:$
$$\frac{1}{4} \frac{d}{dt} \int_{\Omega} v^4\, dx   +3\int_{\Omega} (\nu \rho v_x-p[\eta,\theta])\ v^2v_x\, dx$$
$$=\int_{\Omega} gv^3\, dx -p_{\Gamma}\left. v^3\right|_{x=M}.$$
By summing up equality (\ref{l71}) and the latter one multiplied by a parameter $\delta\geq 1$, we get
$$\frac{1}{2} \frac{d}{dt} \int_{\Omega} \left[\left( \frac{1}{2} v^2+c_V(\theta-\theta_{\Gamma})\right)^2+\frac{\delta}{2}v^4\right]\, dx
+\int_{\Omega} \left[(1+3\delta)\nu\rho v^2v_x^2+c_V \kappa[\eta,\theta]\rho \theta_x^2\right]\, dx $$
$$=-\int_{\Omega} \left(\nu c_V+\kappa[\eta,\theta]\right)\, \rho v v_x\theta_x\, dx $$
$$+\int_{\Omega} \left[p_0[\eta]v_x\left( \frac{1}{2} v^2+c_V(\theta-\theta_{\Gamma})\right)
+p[\eta,\theta]\left( (1+3\delta)v^2v_x+c_V v \theta_x\right)\right]\, dx$$
$$+\int_{\Omega} gv\left( (\frac{1}{2}+\delta)v^2+c_V(\theta-\theta_{\Gamma})\right)\, dx
-\left. p_{\Gamma}\left(v \left( (\frac{1}{2}+\delta)v^2+c_V(\theta-\theta_{\Gamma})\right)\right)\right|_{x=M}$$
$$=: I_1+I_2+I_3+I_4.$$

Let us estimate the summands in the last equality.
First, by using the two-sided bounds
$\underline{\eta}\leq \eta\leq \overline{\eta}$ and
$\underline{\kappa}\leq \kappa\leq \overline{\kappa}$, we deduce
$$K_1^{-1}\left( \delta\|vv_x\|_{\Omega}^2+\|\theta_x\|_{\Omega}^2\right)
\leq \int_{\Omega} \left[(1+3\delta)\nu\rho v^2v_x^2+c_V \kappa[\eta,\theta]\rho \theta_x^2\right]\, dx  ,$$
and
$$ |I_1|\leq K_2\|vv_x\|_{\Omega}\|\theta_x\|_{\Omega}\leq
\frac{K_2^2}{4\varepsilon} \|vv_x\|_{\Omega}^2+\varepsilon\|\theta_x\|_{\Omega}^2,
\ \forall \varepsilon>0.$$
Second, by using the estimates $|p_0[\eta]|\leq K_3$ and
$$|p[\eta,\theta]|=|p[\eta,\theta_{\Gamma}]+p_1[\eta](\theta-\theta_{\Gamma})|\leq K_4\left(1+|\theta-\theta_{\Gamma}|\right),$$
 we have
$$ |I_2|\leq K_5\left[
\int_{\Omega} \left( \delta v^2 |v_x|+|v\theta_x|\right)\, dx
+\int_{\Omega} |\theta-\theta_{\Gamma}| \left(|v_x|+\delta v^2 |v_x|+|v\theta_x|\right)\, dx \right]
=: K_5(I_{21}+I_{22}).$$
Furthermore the following estimates hold, for any $\varepsilon>0$:
$$I_{21}\leq \delta \|vv_x\|_{\Omega}\|v\|_{\Omega}+\|v\|_{\Omega} \|\theta_x\|_{\Omega}
\leq \varepsilon\left(\delta \|vv_x\|_{\Omega}^2+\|\theta_x\|_{\Omega}^2\right)+\frac{\delta+1}{4\varepsilon} \|v\|_{\Omega}^2,$$
and
$$I_{22}
\leq \|\theta-\theta_{\Gamma}\|_{L^{\infty}(\Omega)}\left\| \frac{v_x}{\sqrt{\theta}}\right\|_{\Omega}\|\theta\|_{L^1(\Omega)}^{1/2}
+\|\theta-\theta_{\Gamma}\|_{\Omega}\|v\|_{L^{\infty}(\Omega)} \left(\delta \|vv_x\|_{\Omega}+\|\theta_x\|_{\Omega}\right)$$
$$\leq \varepsilon \left(\frac{\delta}{2} \|vv_x\|_{\Omega}^2+\|\theta_x\|_{\Omega}^2\right)
+\frac{MK^{(1)}}{2\varepsilon} \left\| \frac{v_x}{\sqrt{\theta}}\right\|_{\Omega}^2
+\frac{\delta}{\varepsilon}\|v\|_{L^{\infty}(\Omega)}^2 \|\theta-\theta_{\Gamma}\|_{\Omega}^2.$$
Third, we obtain
$$|I_3|+|I_4|
\leq \left( \|g\|_{L^1(\Omega)}+p_{\Gamma}\right)\,
\|v\|_{C(\overline{\Omega})}M^{1/2}\|(1+2\delta) vv_x+c_V\theta_x\|_{\Omega}$$
$$\leq \varepsilon \left(\delta \|vv_x\|_{\Omega}^2+\|\theta_x\|_{\Omega}^2\right)+\frac{K_6\delta}{\varepsilon} \|v\|_{C(\overline{\Omega})}^2,$$
where all the above quantities $K_i$,$1\leq i\leq 6,$ do not depend on
$\delta $ and $\varepsilon.$

Now, by choosing $\varepsilon:=K_7^{-1}$ small enough and then
$\delta:=K_8 \varepsilon^{-1}$ large enough, and setting
$$y:=\int_{\Omega}\left[\left(\frac{1}{2}v^2+c_V(\theta-\theta_{\Gamma})\right)^2+\frac{\delta}{2}v^4\right]\, dx,$$
we get
\begin{equation}
\frac{dy}{dt}+K_9^{-1} \left( \|vv_x\|_{\Omega}^2+\|\theta_x\|_{\Omega}^2\right)\leq K_{10} (ay+h),
\label{l72}
\end{equation}
with $a:=\|v\|_{L^{\infty}(\Omega)}^2$ and $h:= \left\| \frac{v_x}{\sqrt{\theta}}\right\|_{\Omega}^2$ (see (\ref{c11})); moreover
$$K_{11}^{-1} \left(\frac{1}{2}\|v^2\|_{\Omega}^2+\|\theta-\theta_{\Gamma}\|_{\Omega}^2\right)
\leq y
\leq K_{11} \left(\frac{1}{2}\|v^2\|_{\Omega}^2+\|\theta-\theta_{\Gamma}\|_{\Omega}^2\right).$$
It is clear that
$$\frac{dy}{dt}+K_{12}^{-1} y\leq K_{10} (ay+h),$$
with $K_{12}:= K_9 K_{11} M^2$. By corollary \ref{cor1} we have
$\|a\|_{L^1({\R}^+)}\leq K^{(1)} \|h\|_{L^1({\R}^+)}\leq \left(K^{(5)}\right)^2$, so lemma \ref{lem6} implies
$$y(t)\leq K_{13}\ \mbox{on}\ \overline{{\R}}^+,\  \mbox{and}\ y(t)\rightarrow 0\  \mbox{as}\ t\rightarrow \infty.$$
By integrating inequality (\ref{l72}) over ${\R}^+$, we also obtain
$$K_9^{-1} \left( \|vv_x\|_Q^2+\|\theta_x\|_Q^2\right)
\leq y(0)+K_{10}\left( \|a\|_{L^1({\R}^+)}\ \sup_{\overline{{\R}}^+}y+\|h\|_{L^1({\R}^+)}\right),$$
so that $\|vv_x\|_Q+\|\theta_x\|_Q\leq K_{14}$, and the lemma is proved.$\ \ \ \Box$
\vskip0.2cm
Let us now estimate $v_x$ in $L^2(Q)$.

\newtheorem{lem8}[lem1]{Lemma}
 \begin{lem8}
 The following estimate holds
$$\|v_x\|_Q \leq K^{(7)}.$$
 \label{lem8}
\end{lem8}
{\bf Proof:} By taking $L^2(\Omega)$-inner product of
the second equation (\ref{e1}) with $v$, we get the equality
(compare with (\ref{l14}))
$$ \frac{d}{dt} \int_{\Omega} \left( \frac{1}{2} v^2+p_S \eta- P[\eta,\theta_{\Gamma}] \right)\, dx
  +\int_{\Omega}\nu \rho v_x^2\, dx
=\int_{\Omega} p_1[\eta](\theta-\theta_{\Gamma})\, v_x\, dx.$$
By integrating it over $(0,T)$ and exploiting the bounds
$\underline{\eta}\leq \eta\leq \overline{\eta}$, we get
$$\|v_x\|_{Q_T}^2 \leq K_1 \left( 1+\|\theta-\theta_{\Gamma}\|_{Q_T}\|v_x\|_{Q_T}\right).$$
So $\|v_x\|_{Q_T} \leq K_1^{1/2}+K_1\|\theta-\theta_{\Gamma}\|_{Q_T}
\leq K_1^{1/2}+K_1M\|\theta_x\|_{Q_T},\ \mbox{for any}\ T>0$, and the result
follows from the previous lemma.$\ \ \ \Box$
\vskip0.2cm
Now we prove additional properties of $p[\eta,\theta]-p_S$.

\newtheorem{lem9}[lem1]{Lemma}
 \begin{lem9}
 For $p[\eta,\theta]-p_S$, the following estimate together with stabilization property hold
 \begin{equation}
\|p[\eta,\theta]-p_S\|_Q\leq K^{(8)},
\label{l91}
\end{equation}
\begin{equation}
\|p[\eta,\theta](\cdot,t)-p_S(\cdot)\|_{\Omega} \rightarrow 0\  \mbox{as}\ t\rightarrow \infty.
\label{l92}
\end{equation}
 \label{lem9}
\end{lem9}
{\bf Proof:}

1. Equation (\ref{l30}) implies the following equality, for any $T>0$
 $$\|p[\eta,\theta]-p_S\|_{Q_T}^2 +\|I^*v_t\|_{Q_T}^2
= \|\nu \rho v_x \|_{Q_T}^2+2\int_{Q_T} \left( p[\eta,\theta]-p_S\right)\,
I^*v_t\, dx\, dt.$$
Elementary transformations and the bounds
$\underline{\eta}\leq \eta\leq \overline{\eta}$ give
$$\int_{Q_T} \left( p[\eta,\theta]-p_S\right)\, I^*v_t\, dx\, dt
=\int_{Q_T} \left( p[\eta,\theta_{\Gamma}]-p_S\right)\, I^*v_t\, dx\, dt
+\int_{Q_T}  p_1[\eta](\theta-\theta_{\Gamma})\, I^*v_t\, dx\, dt$$
$$=\left. \int_{\Omega} \left( p[\eta,\theta_{\Gamma}]-p_S\right)\, I^*v\, dx\right|_0^T
-\int_{Q_T}  p_{\eta}[\eta,\theta_{\Gamma}]\ \eta_t\, I^*v\, dx\, dt
+\int_{Q_T}  p_1[\eta](\theta-\theta_{\Gamma})\, I^*v_t\, dx\, dt$$
$$\leq K_1\left( \|v(\cdot,T)\|_{\Omega}+\|v^0\|_{\Omega}+\|v_x\|_{Q_T}\|v\|_{Q_T}+\|\theta-\theta_{\Gamma}\|_{Q_T}\ \|I^*v_t\|_{Q_T}\right).$$
Therefore
$$\|p[\eta,\theta]-p_S\|_{Q_T}^2+\frac{1}{2}\|I^*v_t\|_{Q_T}^2
\leq \nu \underline{\eta}^{-2} \|v_x\|_{Q_T}^2
+ K_1\left( \|v(\cdot,T)\|_{\Omega}+N+M\|v_x\|_{Q_T}^2 \right)+(K_1M)^2\|\theta_x\|_{Q_T}^2 ,$$
so estimate (\ref{l91}) follows from lemmas \ref{lem1}, \ref{lem7}, and \ref{lem8}.

2. First, instead of property (\ref{l92}), let us prove that
\begin{equation}
\|p[\eta,\theta_{\Gamma}](\cdot,t)-p_S(\cdot)\|_{\Omega} \rightarrow 0\  \mbox{as}\ t\rightarrow \infty.
\label{l93}
\end{equation}
By using the estimates $\underline{\eta}\leq \eta$, (\ref{l91}), and $\|\theta_x\|_Q\leq K^{(6)}$, we have
\begin{equation}
\|p[\eta,\theta_{\Gamma}]-p_S\|_Q\leq \|p[\eta,\theta]-p_S\|_Q+\|p_1[\eta]\|_{L^{\infty}(Q)} \|\theta-\theta_{\Gamma}\|_Q\leq K_2.
\label{l94}
\end{equation}
Then also
$$\int_0^{\infty}\left| \frac{d}{dt} \left( \|p[\eta,\theta_{\Gamma}]-p_S\|_{\Omega}^2\right)\right|\, dt
=2\int_0^{\infty}\left|\int_{\Omega}  p_{\eta}[\eta,\theta_{\Gamma}]\eta_t (p[\eta,\theta_{\Gamma}]-p_S)\, dx\right|\, dt$$
\begin{equation}
\leq 2\|p_{\eta}[\eta,\theta_{\Gamma}]\|_{L^{\infty}(Q)}\|v_x\|_Q\|p[\eta,\theta_{\Gamma}]-p_S\|_Q\leq K_3.
\label{l95}
\end{equation}
Estimates (\ref{l94}) and (\ref{l95}) imply property (\ref{l93}).

But by the bounds $\underline{\eta}\leq \eta\leq \overline{\eta}$
the stabilization property (\ref{l70}) we get
$$\left| \|p[\eta,\theta]-p_S\|_{\Omega}^2-\|p[\eta,\theta_{\Gamma}]-p_S\|_{\Omega}^2\right|$$
$$\leq \left[2M^{1/2} \left( \|p[\eta,\theta_{\Gamma}]\|_{L^{\infty}(\Omega)}
+\overline{p}_S\right)
+ \|p_1[\eta]\|_{L^{\infty}(\Omega)}\ \|\theta-\theta_{\Gamma}\|_{\Omega}\right]
\ \|p_1[\eta]\|_{L^{\infty}(\Omega)} \|\theta-\theta_{\Gamma}\|_{\Omega} $$
$$\leq K_4(1+\|\theta-\theta_{\Gamma}\|_{\Omega})
\|\theta-\theta_{\Gamma}\|_{\Omega}
\rightarrow 0\  \mbox{as}\ t\rightarrow \infty,$$
so that (\ref{l93}) implies (\ref{l92}).$\ \ \ \Box$
\vskip0.2cm

To establish the pointwise convergence of the specific volume $\eta(x,t)$ as $t\rightarrow \infty$, we need a modification of the Ball-Pego
lemma \cite{P87} concerning ``almost autonomous" ordinary differential equations.

\newtheorem{lem10}[lem1]{Lemma}
 \begin{lem10}
Let $f\in C({\R})$ be such that, for a given constant $f_S$, there exists no interval $(z_1,z_2)$ such that $f(z)\equiv f_S$ on
$(z_1,z_2)$. Let also $\alpha, \beta\in C({\R}^+)$ be two functions such that
 ${\displaystyle \alpha(t)\rightarrow 0}$ and
${\displaystyle \beta(t)\rightarrow 0\ \
 \mbox{as}\ t\rightarrow \infty}$, as well as
$a\in L^1({\R}^+)$.

 If a function $y$ satisfies ${\displaystyle \sup_{{\R}^+}|y(t)|<\infty}$, $y\in W^{1,1}(0,T)$ for all $T>0$, and

$${\displaystyle \frac{dy}{dt}=f(y+\alpha)-f_S+a+\beta\ \ \mbox{on}\ {\R}^+,}$$
then
$${\displaystyle y(t)\rightarrow y_S\ \ \mbox{as}\ t\rightarrow \infty
,\ \mbox{and}\ f(y_S)=f_S.}$$

The result remains valid if one sets $\beta=0$ and replaces the condition $a\in L^1({\R}^+)$ by the
following ones
$${\displaystyle |a|\leq |a_1|+|\beta_1|,\ \ a,a_1,\beta_1\in C({\R}^+),\ \ a_1\in L^1({\R}^+),
\ \ \mbox{and}\ \beta_1(t) \rightarrow 0\ \ \mbox{as}\ t\rightarrow \infty.}$$
 \label{lem10}
\end{lem10}
{\bf Proof:} We set $A(t):=\int_t^{\infty} a(\tau)\, d\tau$ and, for $z:=y-A$, we get
\begin{equation}
\frac{dz}{dt}=f(z+\widetilde{\alpha})-f_S+\beta,
\label{l101}
\end{equation}
where $\widetilde{\alpha}:=\alpha +A \in C({\R}^+)$ and
${\displaystyle \widetilde{\alpha}(t) \rightarrow 0\ \ \mbox{as}\ t\rightarrow \infty}$.
Note that $z\in C^1({\R}^+)$, in virtue of equation (\ref{l101}), and
$${\displaystyle \sup_{{\R}^+} |z(t)|\leq \sup_{{\R}^+} |y(t)|+\|a\|_{L^1({\R}^+)}.}$$

Suppose that ${\displaystyle z_1:=\liminf_{t\rightarrow \infty}z(t)<z_2:=\limsup_{t\rightarrow \infty}z(t)}$.
Then for any $z_0\in (z_1,z_2)$, there exist two sequences $\{t_{1k}\}$ and $\{t_{2k}\}$ such that
$$z(t_{1k})=z(t_{2k})=z_0,\ \ \frac{dz}{dt}(t_{1k})\geq 0,\ \ \frac{dz}{dt}(t_{2k})\leq 0.$$
Equation (\ref{l101}) applied for $t=t_{1k}$ and $t=t_{2k}$ as $k\rightarrow \infty$ implies that $f(z_0)-f_S=0$. So by contradiction with the
condition on $f$, ${\displaystyle z_1=z_2=z_S:=\lim_{t\rightarrow \infty}z(t)}$.

By integrating equation (\ref{l101}) over the interval $(k-1,k)$ and
passing to the limit as $k\rightarrow \infty$, we obtain: $f(z_S)-f_S=0$.
It remains to use the equality
$\lim_{\, t\rightarrow \infty}y(t)=\lim_{\, t\rightarrow \infty}z(t)$ to obtain the required result.

To prove the last part of the lemma, it suffices to apply the decomposition
$a=\widetilde{a}+\widetilde{\beta}$, with
$\widetilde{a}:= \frac{a}{|a_1|+\widetilde{\beta}_1} |a_1|$,
$\widetilde{\beta}:= \frac{a}{|a_1|+\widetilde{\beta}_1} \widetilde{\beta}_1$
and $\widetilde{\beta}_1(t):=|\beta(t)|+\frac{1}{t+1}$; here $\widetilde{a}\in L^1({\R}^+)$,
$\widetilde{\beta}\in C({\R}^+),$ and
$\widetilde{\beta}(t)\rightarrow 0\ \ \mbox{as}\ t\rightarrow \infty$
(since $|\widetilde{a}|\leq |a_1|$ and
$|\widetilde{\beta}|\leq |\beta(t)|+\frac{1}{t+1}$).$\ \ \ \Box$

\newtheorem{lem11}[lem1]{Lemma}
 \begin{lem11}
Let condition (\ref{t3}) be satisfied. Then the following pointwise stabilization property holds for the specific volume $\eta$:
there exists a function $\eta_S\in L^{\infty}(\Omega)$ satisfying (\ref{t4}) such that
$$\eta(x,t)\rightarrow \eta_S(x)\ \mbox{as}\ t\rightarrow \infty,\
\mbox{for all}\ x\in\overline{\Omega}.$$
\label{lem11}
\end{lem11}
{\bf Proof:} For any fixed $x\in\overline{\Omega}$, we rewrite equation (\ref{l31}) in the following form
\begin{equation}
{\displaystyle \frac{dy}{dt}=f(y+\alpha)-p_S+p_1[\eta](\theta-\theta_{\Gamma}),}
\label{l111}
\end{equation}
with $y:=\nu \log\eta-\alpha$, $\alpha:=-I^*v$, and $f(z):= p\left( \exp(\frac{z}{\nu}),\theta_{\Gamma}\right)$.
Property (\ref{t3}) yields the corresponding property of $f$ in lemma \ref{lem10}, for any $f_S=p_S(\cdot)$.

By using the bounds $\underline{\eta}\leq \eta\leq \overline{\eta}$ and  the stabilization property (\ref{l70}) we get
$${\displaystyle \sup_{t\geq 0} |y(t)|
\leq \nu\, \max \{ |\log \underline{\eta}|,|\log \overline{\eta}|\}+M^{1/2} \|v\|_{L^{2,\infty}(Q)}\leq K_1,}$$
$$|\alpha(t)|\leq M^{1/2} \|v(\cdot,t)\|_{\Omega}\rightarrow 0\ \ \mbox{as}\ t\rightarrow \infty.$$
We also have, by the H\"older inequality for numbers
$$\left| p_1[\eta](\theta-\theta_{\Gamma})\right|\leq \|p_1[\eta]\|_{L^{\infty}(Q)}\|\theta-\theta_{\Gamma}\|_{C(\overline{\Omega})}
\leq K_2 \|\theta_x\|_{\Omega}^{1/2}\|\theta-\theta_{\Gamma}\|_{\Omega}^{1/2}$$
$$\leq \|\theta_x\|_{\Omega}^2+K_2^{4/3}\|\theta-\theta_{\Gamma}\|_{\Omega}^{2/3}=: a_1+\beta_1.$$
The functions
$a(\cdot,t):=p_1[\eta](\cdot,t)(\theta(\cdot,t)-\theta_{\Gamma})$ and $a_1$, $\beta_1$ satisfy the conditions of the final part of lemme \ref{lem10}
by virtue of lemma \ref{lem7}
(together with the properties
$\eta(\cdot,t),\theta(\cdot,t), \|\theta_x(\cdot,t)\|_{\Omega}\in C({\R}^+))$.

So by condition (\ref{t3}) and lemma \ref{lem10}, there exists
$${\displaystyle \lim_{t\rightarrow \infty} y(t)=y_S,\ \mbox{with}\ f(y_S)=p_S,}$$
i.e. $\eta(\cdot,t)\rightarrow \eta_S(\cdot)=\exp(\frac{y_S}{\nu})\ \mbox{as}\ t\rightarrow \infty$ and
$p(\eta_S(\cdot),\theta_{\Gamma})=p_S(\cdot)$.
The bounds
$\underline{\eta}\leq \eta\leq \overline{\eta}$ and the measurability of $\eta(\cdot,t)$ on $\Omega$ imply the bounds
$\underline{\eta}\leq \eta_S\leq \overline{\eta}$ and the measurability of $\eta_S$ on $\Omega$.$\ \ \ \Box$
\vskip0.2cm
Note that the Lebesgue dominated convergence theorem immediately gives
$$\|\eta(\cdot,t)-\eta_S(\cdot)\|_{L^q(\Omega)}\rightarrow 0\ \mbox{as}\ t\rightarrow \infty,\ \mbox{for any}\ q\in[1,\infty).$$
\vskip0.2cm
To prove the stabilization for $v$ in $L^q(\Omega)$, we turn
to the auxiliary linear parabolic problem
\begin{equation}
\left\{  \begin{array}{ll}
u_t=(\mu u_x-\phi)_x+g\ \mbox{in}\ Q, \\[3mm]
\left.             u \right|_{x=0}=0,\, \
\left. (\mu u_x-\phi)\right|_{x=M}=-p_{\Gamma}(t),\, \
\left.              u\right|_{t=0}=u^0(x).\\
  \end{array}
\right.
\label{l120}
\end{equation}
Suppose that $\mu\in L^{\infty}(Q_T)$ and $\mu_t\in L^2(Q_T)$ for any $T>0$, with
$0<\underline{\mu}\leq \mu$ in $Q$. Suppose also that
$\phi\in L^{2,\infty}(Q)$,
$g\in L^{1,\infty}(Q)$, $p_{\Gamma}\in L^{\infty}({\R}^+)$, and that
$u^0\in H^1(\Omega)$, with $u^0(0)=0$.

Set $|\|u|\|_q:=\|u\|_{L^{q,\infty}(Q)}+\|u\|_{L^{\infty,q}(Q)}$
to shorten the notation.

\newtheorem{lem12}[lem1]{Lemma}
 \begin{lem12}
Let $u\in H^1(Q_T)\cap L^{\infty}(Q_T)$ for any $T>0$ be a weak solution to problem (\ref{l120}) such that $|\|u|\|_2<\infty$.
Then, for any $q\in [2,\infty),$
the following estimate together with stabilization property hold
$$|\|u|\|_q\leq C\left[ \|u^0\|_{L^{q}(\Omega)}+q\left( \|\phi\|_{L^{2,\infty}(Q)}+\|g\|_{L^{1,\infty}(Q)}
+\|p_{\Gamma}\|_{L^{\infty}({\R}^+)}+|\|u|\|_2\right)\right],$$
$$\|u(\cdot,t)\|_{L^{q}(\Omega)} \rightarrow  0\ \mbox{as}\ t\rightarrow \infty,$$
where $C$ depends only on $\underline{\mu}$ and $M$.

\label{lem12}
\end{lem12}

More general assertions of such kind (together with applications to barotropic fluid equations) were given in \cite{Z92}, \cite{ZB94},
\cite{Z00}, and the lemma follows from these assertions.

\newtheorem{lem13}[lem1]{Lemma}
 \begin{lem13}
Let $\|v^0\|_{L^{q}(\Omega)}\leq N$, for some $q\in (4,\infty)$.
For $v$, the following estimate together with stabilization property hold
$$|\|v|\|_q\leq q K^{(9)},$$
$$\|v(\cdot,t)\|_{L^{q}(\Omega)} \rightarrow  0\ \mbox{as}\ t\rightarrow \infty,$$
where $K^{(9)}$ is independent of $q$.
\label{lem13}
\end{lem13}
{\bf Proof:} We consider $v$ as the solution to problem (\ref{l120}) with given $\mu:=\nu \rho$, $\phi:=p[\eta,\theta]$.
 By the bounds $\underline{\eta}\leq \eta\leq \overline{\eta}$ and
lemma \ref{lem7}, the following estimates are valid
 $$K_1^{-1}\leq \mu,$$
$$\|\phi\|_{L^{2,\infty}(Q)}
\leq M^{1/2} \|p[\eta,\theta_{\Gamma}]\|_{L^{\infty}(Q)}+\|p_1[\eta]\|_{L^{\infty}(Q)}\ \|\theta-\theta_{\Gamma}\|_{L^{2,\infty}(Q)}
\leq K_2,$$
$$|\|v|\|_2\leq K_3,$$
and the result is proved, by appling the previous lemma \ref{lem12}.$\ \ \ \Box$
\vskip0.5cm
By collecting all of the results of the above lemmas the proof of theorem \ref{th1} is complete.$\ \ \ \Box$

\subsection{Proof of proposition \ref{prop1}}

Note that condition $N^{-1}\leq \underline{p}_{\, S}$ has been used above in lemma \ref{lem1}, but not in lemma \ref{lem3}.

Let us turn to the proof of lemma \ref{lem1}, supposing that in contrast to (\ref{pr11}), we have
\begin{equation}
\overline{V}:= \sup_{t\geq 0} V(t)<\infty.
\label{l141}
\end{equation}
By using the formula
$p_S\eta=\varepsilon \eta+(p_S-\varepsilon)\eta$ and the estimate
$$\left| \int_{\Omega} (p_S-\varepsilon)\eta\, dx\right|\leq \left( |p_{\Gamma}|
+\|g\|_{L^1(\Omega)}+\varepsilon \right)\, \overline{V}\ \
\forall \varepsilon>0,$$
we see that lemma \ref{lem1} remains valid and consequently lemma
\ref{lem3} is also valid.
The quantities $K^{(1)}-K^{(3)}$ now depend on $\overline{V}$ as well.

Consider equation (\ref{l111}). By applying the operator $I_0$ to it and
exploiting the bound
$\underline{\eta}\leq \eta$, we get
$$\nu \log \eta\geq \nu \log \eta^0-I^*(v-v_0)
+I_0(p[\eta,\theta_{\Gamma}]-p_S)-K_1 I_0 \max\{ \theta_{\Gamma}-\theta, 0\} $$
as $p_1[\eta ] \leq K_1.$
Let us introduce the set $E_t:=\{ x\in \overline{\Omega}\ :\
\theta(x,t)\leq \theta_{\Gamma}\}$. Then
$$\|\max\{ \theta_{\Gamma}-\theta(\cdot,t), 0\}\|_{C(\overline{\Omega})}
\leq \| \theta_x(\cdot,t)\|_{L^1(E_t)}
\leq \| \frac{\theta_{\Gamma}}{\theta}\theta_x(\cdot,t)\|_{L^1(E_t)}$$
$$\leq \| \frac{\theta_{\Gamma}}{\theta}\theta_x(\cdot,t)\|_{L^1(\Omega)}
\leq \theta_{\Gamma}V^{1/2}
\| \frac{\sqrt{ \rho}}{ \theta}\theta_x(\cdot,t)\|_{\Omega}.$$
By using estimates (\ref{l12}) and (\ref{l141})
$$I_0\max\{ \theta_{\Gamma}-\theta(\cdot,t), 0\}
\leq \theta_{\Gamma} {\overline{V}}^{1/2}K^{(2)}t^{1/2}.$$
This estimate together with (\ref{l32}) imply
\begin{equation}
\nu \log \eta \geq -\frac{1}{\varepsilon}K_2-\varepsilon t
+I_0(p[\eta,\theta_{\Gamma}]-p_S),\ \ \forall \varepsilon\in (0,1).
\label{l142}
\end{equation}
As $\underline{p}_{\, S}<m(\theta_{\Gamma})$, for some $x_0$ and
for $\varepsilon_0>0$ and $\delta>0,$ both small enough, we have
$$p_S(x)\leq m(\theta_{\Gamma})-\varepsilon_0,\ \
\mbox{for}\ x\in [x_0,x_0+\delta]\subset \overline{\Omega}.$$
By choosing $\varepsilon:=\varepsilon_0/2$, estimate (\ref{l142}) gives
$$\nu \log \eta \geq \frac{1}{2}\varepsilon_0 t-\frac{2}{\varepsilon_0} K_2
\ \ \mbox{on}\ [x_0,x_0+\delta]\times \overline{{\R}}^+.$$
But then
$$V(t)\geq K_3\delta \exp\left(\frac{\varepsilon_0}{2\nu} t\right) \rightarrow  \infty\ \mbox{as}\ t\rightarrow \infty,$$
with $K_3:=\exp\left(-\frac{2}{\nu \varepsilon_0} K_2\right)$, which clearly contradicts (\ref{l141}).$\ \ \ \Box$

\subsection{Proof of proposition \ref{prop2}}

Suppose that in contrast to (\ref{pr21})
\begin{equation}
\sup_{t\geq 0}\left|\int_{\Omega} v(x,t)\, dx\right|\leq C_1<\infty.
\label{l150}
\end{equation}
Set $\eta_0(t):=\eta(0,t),$ consider equation (\ref{l31}) for $x=0$ and integrate it in $t$:
\begin{equation}
\nu \log \eta_0(t)= \nu \log \eta^0(0)
+\int_{\Omega} (v^0(x)-v(x,t))\, dx
+\int_0^t p(\eta_0(\tau),\theta_{\Gamma})\, d\tau
\label{l151}
\end{equation}
as $\theta |_{x=0}=\theta_{\Gamma}$ and $p_S(0)=0.$ It is straightforward that
(see (\ref{t1}) and (\ref{l150}))
\begin{equation}
\left|\nu \log \eta^0(0)+\int_{\Omega} (v^0(x)-v(x,t))\, dx\right|\leq K_1+C_1.
\label{l152}
\end{equation}

Now set $b(t):=\int_0^tp(\eta_0(\tau),\theta_{\Gamma})\, d\tau$.
As $p(\eta ,\theta_{\Gamma})>m(\theta_{\Gamma})=0,$
the function $b$ is increasing and positive on ${\R}^+$.
Let us show the property
\begin{equation}
b(t) \rightarrow  \infty\ \mbox{as}\ t \rightarrow \infty.
\label{l153}
\end{equation}
Indeed if, in contrast to this property, $0<b(t)\leq C_2$ on ${\R}^+$,
then according to (\ref{l151}) and (\ref{l152})
$$0<\eta_0(t)\leq C_3\ \ \mbox{on}\ \overline{{\R}}^+.$$
This estimate implies
$p(\eta_0(t),\theta_{\Gamma})\geq \varepsilon_0>0$ on $\overline{{\R}}^+$
and so $b(t)\geq \varepsilon_0 t$ on ${\R}^+$.
This contradiction proves (\ref{l153}).

Property (\ref{pr22}) immediately follows from (\ref{l151})-(\ref{l153}).

Let us justify the last part of proposition \ref{prop2}.
By the conditions on $p_S$ and $p(\eta, \theta_{\Gamma})$, we can consider
$$0\leq p_S \eta,\ \ -P(\eta, \theta_{\Gamma})=\int_{\eta}^{\infty} p(\zeta, \theta_{\Gamma})\ d\zeta>0.$$
So if we turn to the proof of lemma \ref{lem1}, we see that it remains valid
but only the first summand in (\ref{l11}) should be dropped.
 In particular $\|v\|_{L^{2,\infty}(Q)}\leq K^{(1)}$, consequently property
(\ref{l150}) holds, and by the first part of the proof
so does property (\ref{pr22}).

\subsection{Proof of theorem \ref{th2} }

Properties (\ref{p2}) and (\ref{p3}) imply the following estimates
$$-(P(\eta,\theta_{\Gamma})-P(\check{\eta},\theta_{\Gamma}))
=\int^{\check{\eta}}_{\eta}(p_0(\zeta)+p_1(\zeta)\, \theta_{\Gamma})\, d\zeta
\geq \overline{p}_S(\check{\eta}-{\eta})
\ \ \mbox{for}\ 0<\eta\leq \check{\eta},$$
$$P(\eta,\theta_{\Gamma})-P(\hat{\eta},\theta_{\Gamma})
=\int_{\hat{\eta}}^{\eta}(p_0(\zeta)+p_1(\zeta)\, \theta_{\Gamma})\, d\zeta
\leq \underline{p}_{\, S}(\hat{\eta}-{\eta})
\ \ \mbox{for}\ \hat{\eta} \leq \eta.$$
Therefore
$$p_S \eta -P(\eta,\theta_{\Gamma})\geq
C:=\mbox{min}\{ -P(\check{\eta},\theta_{\Gamma})+\overline{p}_S\check{\eta},
\, -\mbox{max}_{ \, \check{\eta}
\leq \eta \leq \hat{\eta}} P(\eta,\theta_{\Gamma}),
\, -P(\hat{\eta},\theta_{\Gamma})+\underline{p}_{\, S}\hat{\eta} \}
\ \ \mbox{for all}\ \eta>0.$$
This means that lemma \ref{lem1} remains valid but only the first summand
in (\ref{l11}) should be dropped.

In order to check the bounds in lemmas \ref{lem3} and \ref{lem5}, we can use
the properties, respectively
$$p(\eta,\theta)-p_S(x)\geq 0\ \ \mbox{for}\ 0<\eta\leq \check{\eta},\
0< \theta, \ \mbox{and}\ x\in\overline{\Omega},$$
$$p(\eta,\theta)-p_S(x)\leq 0\ \ \mbox{for}\ \hat{\eta}\leq \eta,\
0< \theta, \ \mbox{and}\ x\in\overline{\Omega}$$
(see properties (\ref{p2}) and (\ref{p3})). But by using equation (\ref{l31}),
estimate (\ref{l32}), and the remark after lemma \ref{lem2}
(with $N_1=0$ and $\varepsilon_0=0$), the uniform bounds
$\underline{\eta}\leq \eta(x,t)$ and $\eta(x,t)\leq \overline{\eta}$
in $Q$ hold.

After the bounds $\underline{\eta}\leq \eta \leq \overline{\eta}$, in fact,
the rest of the proof of theorem \ref{th1} remains unchanged.

 \vskip0.2cm
{\bf Acknowledgments:}
 \vskip0.2cm

This work was mainly accomplished while the second author was visiting
the Department of Mathematics of the University Paris 7-Denis Diderot
which he thanks for hospitality. He was also partially supported by RFBR
projects 00-01-00207 and 01-01-00700.

\section*{Appendix}

This appendix is devoted to the proof of the existence of a
regular weak solution to the problem (\ref{e1}),(\ref{e7}), and (\ref{e8}).

\newtheorem{prop3}[prop1]{Proposition}
 \begin{prop3}
 Suppose that either conditions (\ref{p0}), (\ref{p1}), and
$N^{-1}\leq \underline{p}_{\, S}$,
or (\ref{p2}) and (\ref{p3}) are valid. Suppose also that
 $\kappa_{\eta \eta}\in C({\R}^+\times{\R}^+)$ and
$\eta^0,v^0,\theta^0\in H^1(\Omega)$, $g\in L^2(\Omega)$ with
 $$\|\eta^0\|_{H^1(\Omega)}+\|v^0\|_{H^1(\Omega)}
+\|\theta^0\|_{H^1(\Omega)}\leq N,\ \ \|g\|_{L^2(\Omega)}\leq N,$$
 $$N^{-1}\leq \eta^0,\ \ N^{-1}\leq \theta^0,\ \ v^0(0)=0,
\ \ \theta^0(0)=\theta_{\Gamma}.$$
Then for any $T>0$, the problem (\ref{e1}),(\ref{e7}), and (\ref{e8})
admits a unique regular weak solution, and it satisfies the following
estimates
\begin{equation}
\|\eta_x\|_{L^{2,\infty}(Q_T)}+\|\eta_t\|_{L^{2,\infty}(Q_T)}
+\|v\|_{H^{2,1}(Q_T)}+\|\theta\|_{H^{2,1}(Q_T)}\leq K^{(10)},
\label{a1}
\end{equation}
\begin{equation}
0<\underline{\eta}\leq \eta(x,t) \leq \overline{\eta},
\ \ 0<\underline{\theta}:=(K^{(11)})^{-1}\leq \theta(x,t)
\ \mbox{in}\ \overline{Q}_T.
\label{a2}
\end{equation}
Hereafter, the quantities $K_i$ and $K^{(i)}$ may depend also on $T$.
\label{prop3}
\end{prop3}

\vskip0.2cm
{\bf Proof:} We shall exploit a priori estimates given in
theorems \ref{th1} and \ref{th2} and derive additional
estimates in $Q_T$ in several steps. We shall finish by the proof of a local
(in time) existence theorem.

1. We set $w:=\nu (\log \eta)_x-v$ and
 rewrite the second equation (\ref{e1}) as follows
 $$w_t=\left( p_{0\eta}[\eta]+p_{1\eta}[\eta]\theta\right) \eta_x
+p_1[\eta]\theta_x-g.$$
 By taking $L^2(\Omega)$-inner product with $w$, using the formula
 $\eta_x=\frac{1}{\nu}\eta(w+v)$ and the bounds
$\underline{\eta} \leq \eta\leq \overline{\eta}$, we obtain the inequality
 $$\frac{d}{dt} \|w\|_{\Omega}^2
\leq K_1\left[ (1+\|\theta\|_{L^{\infty}(\Omega)})(\|w\|_{\Omega}^2
+\|v\|_{\Omega}^2)+\|\theta_x\|^2_{\Omega}+\|g\|^2_{\Omega}\right].$$
The estimates
$\|\theta\|_{L^{\infty}(\Omega)}
\leq \theta_{\Gamma}+\sqrt{M}\|\theta_x\|_{\Omega}$,
$\| \theta_x \|_Q \leq K^{(6)}$, and
$\|\nu(\log \eta^0)_x-v^0\|_{\Omega}\leq K_2$,
together with the Gronwall lemma imply the bound
$\|w\|_{L^{2,\infty}(Q_T)}\leq K_3$
and therefore
\begin{equation}
\|\eta_x\|_{L^{2,\infty}(Q_T)}\leq K^{(12)}.
\label{a3}
\end{equation}
Consequently, the function $\rho$ is a H\"{o}lder continuous one
on $\overline Q_T.$

2. The function $u:=I^*v$ satisfies the nondivergent parabolic
problem (see (\ref{l30}) and (\ref{e7}),(\ref{e8}))
\begin{equation}
\left\{  \begin{array}{ll}
u_t=\nu \rho u_{xx}+p[\eta, \theta]-p_S\ \mbox{in}\ Q, \\[3mm]
\left.             u_x \right|_{x=0}=0,\, \ \left.
u\right|_{x=M}=0,\, \
\left.              u\right|_{t=0}=I^*v^0(x).\\
  \end{array}
\right.
\end{equation}
The standard parabolic $H^{2,1;q}(Q_T)-$estimates \cite{LSU}
together with the bounds $\underline{\eta} \leq \eta\leq
\overline{\eta},$ $\|\theta \|_{L^6(Q_T)} \leq c\|\theta
\|_{V_2(Q_T)} \leq K_1$ lead to the estimate
\begin{equation}
\|v_x\|_{L^6(Q_T)}=\|u_{xx}\|_{L^6(Q_T)} \leq K_2\left(\|p[\eta,
\theta]-p_S \|_{L^6(Q_T)} +\|v^0\|_{L^6(\Omega)}\right) \leq
K^{(13)}. \label{a2a}
\end{equation}

3. We also can consider the second equation (\ref{e1}) as a linear
parabolic equation
$$v_t=\left(\nu \rho v_x-p[\eta,\theta]\right)_x+g,$$
with corresponding boundary and initial conditions
(see (\ref{e7}) and (\ref{e8})).
After the bounds $\underline{\eta}\leq \eta\leq \overline{\eta},$
 (\ref{a3}), and (\ref{a2a}), we have $\|\rho_x\|_{L^{2,\infty}(Q_T)}\leq K_1$ and
$$\|p[\eta,\theta]_x\|_{Q_T}\leq K_2\left[ (1+\|\theta\|_{L^{\infty,2}(Q_T)})
\|\eta_x\|_{L^{2,\infty}(Q_T)}+\|\theta_x\|_{Q_T}\right]\leq K_3,$$
$$\|p[\eta,\theta]_t\|_{Q_T}\leq K_4\left[ (1+\|\theta\|_{L^4(Q_T)})
\|v_x\|_{L^4(Q_T)}+\|\theta_t\|_{Q_T}\right]
\leq K_5(1+\|\theta_t\|_{Q_T}).$$
So the standard parabolic $H^{2,1}(Q_T)$-estimates \cite{LSU} (or \cite{AZ97})
imply
\begin{equation}
\|v\|_{H^{2,1}(Q_T)}\leq K_6 \left( \|p[\eta,\theta]\|_{H^1{(Q_T)}}
+\|g\|_{\Omega}
+|p_{\Gamma}|+\|v^0\|_{H^1(\Omega)}\right)
\leq K_7(1+\|\theta_t\|_{Q_T}).
\label{a4}
\end{equation}

4. Let us turn to estimates for $\theta$.
We rewrite equation (\ref{l13a}) as a linear parabolic equation
\begin{equation}
c_{V}\theta_t=\left(A\theta_x\right)_x+F,
\label{a5}
\end{equation}
with $A:=\kappa[\eta,\theta]\rho$ and
$F:=\left( \nu \rho v_x-p_1[\eta]\theta \right) v_x$. By the bounds
$\underline{\kappa}\leq \kappa\leq \overline{\kappa}$ and
$\underline{\eta}\leq \eta\leq \overline{\eta}$, we get
$K_1^{-1}\leq A\leq K_1$ and
\begin{equation}
\|F\|_{Q_T}\leq K_2\left( \|v_x\|_{L^4(Q_T)}
+\|\theta\|_{L^4(Q_T)}\right)\|v_x\|_{L^4(Q_T)}\leq K_3,
\label{a6}
\end{equation}
where the estimates $\|\theta\|_{L^4(Q)}\leq K_4$ and (\ref{a2a})
are again taken into account. Now, the standard parabolic
$L^{\infty}(Q_T)$-estimates \cite{LSU} (or \cite{AZ96}) imply
\begin{equation}
\|\theta\|_{L^{\infty}(Q_T)}
\leq K_5\left(\|F\|_{Q_T}
+\theta_{\Gamma}+\|\theta^0\|_{L^{\infty}(\Omega)}\right)\leq K^{(14)}.
\label{a7}
\end{equation}

5. Let us derive a uniform lower bound for $\theta$.
We divide equation (\ref{a5}) by $-\theta^2$ and transform it as follows
\begin{equation}
c_{V}(\theta^{-1})_t
=\left(A(\theta^{-1})_x\right)_x-2A\theta^{-3}\theta_x^2
-\left( \sqrt{\nu \rho} v_x \theta^{-1}
-\frac{1}{2}\sqrt{\frac{\eta}{\nu}}p_1[\eta]\right)^2
+\frac{\eta}{4\nu} (p_1[\eta])^2.
\label{a8}
\end{equation}
Set $d:=\max\{\theta^{-1}-\theta^{-1}_{\Gamma},0\}$ and note that
$\left. d\right|_{x=0}=0$ and $\left. A(\theta^{-1})_x\right|_{x=M}=0$.
Now we multiply equation (\ref{a8}) by $qd^{q-1}$ with $q\geq 2$,
integrate the result over $\Omega$, apply the bounds
$\underline{\eta}\leq \eta\leq \overline{\eta}$ and the H\"older inequality and obtain
$$c_{V}\frac{d}{dt} \int_{\Omega} d^q\, dx
\leq q \int_{\Omega}\frac{\eta}{4\nu} (p_1[\eta])^2 d^{q-1}\, dx\leq q K_1
\left( \int_{\Omega}d^q\, dx\right)^{\frac{q-1}{q}}.$$
By solving this differential inequality (for example see lemma 1.4
in \cite{Z00}), we find
$$\|d(\cdot,t)\|_{L^q(\Omega)}\leq \|d^0\|_{L^q(\Omega)}+K_1 t,$$
with $d^0:=\max\{{(\theta^0)}^{-1}-\theta^{-1}_{\Gamma},0\}\leq N$.
By passing to the limit as $q\rightarrow \infty$, we get
$$\|d\|_{L^{\infty}(Q_T)}\leq N+K_1T=:K_2.$$
This estimate together with $\theta^{-1}\leq d+\theta^{-1}_{\Gamma}$ imply
\begin{equation}
\underline{\theta}:=(K^{(11)})^{-1}\leq \theta\ \ \mbox{in}\ Q_T.
\label{a9}
\end{equation}

6. Let us prove $H^{2,1}(Q_T)$-bound for $\theta$.
Introduce the function
${\K}(\eta,\theta):=\int_{\theta_{\Gamma}}^{\theta}
\frac{\kappa(\eta,\widetilde{\theta})}{\eta} \ d\widetilde{\theta}$,
and notice that $\left. {\K}[\eta,\theta]\right|_{x=0}=0$.
By taking $L^2(Q_{\tau})$-inner product of equation (\ref{a5})
with ${\K}[\eta,\theta]_t$ we obtain (compare with \cite{KAW})
\begin{equation}
\int_{Q_{\tau}} \left( c_{V} \theta_t {\K}[\eta,\theta]_t
+\pi {\K}[\eta,\theta]_{xt}\right)\, dx\, dt=
\int_{Q_{\tau}} F\cdot {\K}[\eta,\theta]_t\, dx\, dt,\ \ \mbox{for}\ 0
\leq \tau \leq T.
\label{a10}
\end{equation}
The following formulas hold
$${\K}[\eta,\theta]_t={\K}[\eta,\theta]_\eta\ v_x+A\theta_t,\ \
{\K}[\eta ,\theta]_x={\K}_{\eta}[\eta,\theta]\ \eta_x+\pi,$$
$${\K}[\eta,\theta]_{xt}
=\left({\K}_{\eta \eta}[\eta,\theta]\ v_x+{\K}_{\eta \theta}[\eta,\theta]\ \theta_t\right)\eta_x
+{\K}_{\eta }[\eta,\theta]\ v_{xx}+\pi_t.$$
By using the bounds $\underline{\eta}\leq \eta\leq \overline{\eta}$
together with $\underline{\theta } \leq \theta \leq K^{(14)}$
(see (\ref{a7}) and (\ref{a9})) we have
$$|{\K}_{\eta}[\eta,\theta]|+|{\K}_{\eta \eta}[\eta,\theta]|
+|{\K}_{\eta \theta}[\eta,\theta]|\leq K_0.$$
Now from equality (\ref{a10}) it follows that
$$K_1^{-1} \|\theta_t\|_{Q_{\tau}}^2
+\frac{1}{2} \left. \|\pi\|_{\Omega}^2\right|_0^{\tau}$$
$$\leq K_2\int_{Q_{\tau}} \left[ |\theta_t|\, |v_x|
+|\pi|\left((|v_x|+|\theta_t|)|\eta_x|+|v_{xx}|\right)
+|F|(|v_x|+|\theta_t|)\right]\, dx\, dt$$
$$\leq K_2\left[ \|\theta_t\|_{Q_T} \|v_x\|_{Q_T}
+\|\pi\|_{L^{\infty,2}(Q_T)}\left( \|v_x\|_{Q_T}
+\|\theta_t\|_{Q_T} \right) \|\eta_x\|_{L^{2,\infty}(Q_T)} \right.$$
$$\left. +\|\pi\|_{Q_T}\|v_{xx}\|_{Q_T}
+\|F\|_{Q_T}\left( \|v_x\|_{Q_T}+\|\theta_t\|_{Q_T}\right)\right].$$
Let us use the estimates $\|v_x\|_{Q_T}\leq K^{(6)},$
$\|\pi \|_{Q_T} \leq K_3$
as well as (\ref{a3}),(\ref{a4}), and (\ref{a6}), for $\eta_x,v,$ and $F.$
By applying also the estimate
$\|\pi\|_{L^{\infty,2}(Q_T)}\leq \sqrt{2} \|\pi\|^{1/2}_{Q_T}\|\pi_x\|^{1/2}_{Q_T}
\leq \sqrt{2K_3}\|\pi_x\|^{1/2}_{Q_T}$, we get
$$\|\theta_t\|_{Q_T}^2+\|\pi\|_{L^{2,\infty}(Q_T)}^2
\leq K_4\left( 1+\|\pi_x\|_{Q_T}+\|\theta_t \|_{Q_T} \right).$$
By combining this estimate and the trivial one $\|\pi_x\|_{Q_T}\leq c_{V}\|\theta_t\|_{Q_T}+\|F\|_{Q_T}$ (see (\ref{a5})), we obtain
$$\|\theta_t\|_{Q_T}+\|\pi\|_{V_2(Q_T)}\leq K_5.$$
In particular $\|\theta_x\|_{L^{2,\infty}(Q_T)}\leq K_6$ and
$\|\pi\|_{L^{\infty,2}(Q_T)}\leq \sqrt{M} K_5$.

Therefore by using the formula
$$\theta_{xx}=(A^{-1}\pi)_x
=\left( \widetilde{\kappa}_{\eta}[\eta,\theta]\eta_x+ \widetilde{\kappa}_{\theta}[\eta,\theta]\theta_x\right)\pi
+\widetilde{\kappa}[\eta,\theta]\pi_x,$$
with $\widetilde{\kappa}(\eta,\theta):=\frac{\eta}{\kappa(\eta,\theta)}$, we also get
$$\|\theta_{xx}\|_{Q_T}\leq K_7\left[
\left(
\|\eta_x\|_{L^{2,\infty}(Q_T)}+\|\theta_x\|_{L^{2,\infty}(Q_T)}\right)
\|\pi\|_{L^{\infty,2}(Q_T)}+\|\pi_x\|_{Q_T} \right]\leq K_8.$$ So
the estimate $\|\theta\|_{H^{2,1}(Q_T)}\leq K^{(15)}$ is proved.
As a consequence  $\|v\|_{H^{2,1}(Q_T)}\leq K^{(16)}$ (see
(\ref{a4})). This completes the proof of all the a priori
estimates (\ref{a1}) and (\ref{a2}).

 It is not difficult to verify the uniqueness of a regular weak solution
similarly to \cite{AKM}.

7. Now we briefly describe the proof of a local existence theorem. Let us fix
the data satisfying the hypotheses and the additional conditions
\begin{equation}
p_{0\, \eta \eta},p_{1\, \eta \eta}\in C({\R}^+),\ \
\eta^0_{xx}, g_x \in L^2(\Omega).
\label{al1}
\end{equation}
We define the Banach space ${\bf B}_\tau,0<\tau\leq T,$ of triples
$z=(\eta,v,\theta)$ equipped with the norm
$\|z\|_{B_\tau}=\|z\|_{Q_\tau}+\|z_x\|_{L^4(Q_\tau)}+\|\eta_t\|_{Q_\tau}$
and the bounded closed convex set
$$S_\tau=\left \{ z\in {\bf B}_\tau|\,\|z_x\|_{L^4(Q_\tau)}+\|\eta_t\|_{Q_\tau}
\leq N_1,\,
(2N)^{-1}\leq \eta \leq 2c_0N,\, (2N)^{-1}\leq \theta \leq 2c_0N,\, v|_{x=0}=0
\right \},$$
where $N_1>0$ and $c_0$ is such that $\eta^0\leq c_0N,\, \theta^0\leq c_0N.$

We introduce also the nonlinear operator $A:S_\tau \to {\bf B}_\tau$ such that
$A(\widetilde \eta,\widetilde v, \widetilde \theta)=(\eta,v,\theta),$ where
$\theta$ and $v$ satisfy the linear parabolic equations
\begin{equation}
c_V\theta_t=(\kappa[\widetilde \eta,\widetilde \theta]\widetilde \rho \theta_x)_x
+(\nu \widetilde \rho\widetilde v_x
-p_1[\widetilde \eta]\widetilde \theta )\widetilde v_x\ \
\mbox{in}\ Q_\tau,
\label{al2}
\end{equation}
\begin{equation}
v_t=(\nu \widetilde \rho v_x
-p[\widetilde \eta,\theta] )_x+g\ \
\mbox{in}\ Q_\tau,
\label{al3}
\end{equation}
with ${\widetilde \rho}={\widetilde \eta}^{-1},$ and $\eta>0$ satisfies
the ordinary differential equation
\begin{equation}
(\nu \log \eta)_t=p[\eta,\theta]-p_S-I^*v_t\ \
\mbox{in}\ Q_\tau,
\label{al4}
\end{equation}
together with the boundary conditions
\begin{equation}
\theta |_{x=0}=\theta_\Gamma,\
(\kappa[\widetilde \eta,\widetilde \theta]\widetilde \rho \theta_x)
|_{x=M}=0,
\label{al5}
\end{equation}
\begin{equation}
v|_{x=0}=0,\
(\nu \widetilde \rho v_x-p[\widetilde \eta,\theta])|_{x=M}=-p_\Gamma ,
\label{al6}
\end{equation}
and the initial conditions (\ref{e8}).

Problems (\ref{al2}) and (\ref{al5}); (\ref{al3}) and (\ref{al6});
and (\ref{al4}), with the initial conditions (\ref{e8}), can be
solved sequentially. By the linear parabolic equation theory there
exist unique solutions $\theta,v \in H^{2,1}(Q_\tau)$ to the first
and second problems, and they satisfy the estimates
\begin{equation}
\|\theta\|_{H^{2,1}(Q_\tau)}
\leq K_1\exp\left(K_2
\|(\kappa[\widetilde \eta,\widetilde \theta]\widetilde \rho  )_x\|^4_{L^4(Q_\tau)}\right)
\left(1+\|{\widetilde v}_x\|^2_{L^4(Q_\tau)}\right)\leq K_3,
\label{al6a}
\end{equation}
\begin{equation}
\|v\|_{H^{2,1}(Q_\tau)}
\leq K_4\exp\left(K_5(1+\|{\widetilde \rho}_x\|^4_{L^4(Q_\tau)})\right)
\left(1+\|p[\widetilde \eta,\theta]\|_{H^1(Q_\tau)}\right)\leq K_6,
\label{al6b}
\end{equation}
compare with above items 3 and 6. Hereafter the quantities $K_i$ (excluding
$K_1,K_2$ and $K_4,K_5$) depend also on $N_1$.

The following inequalities hold
\begin{equation}
\|\varphi \|_{L^4(Q_\tau)}\leq c_1(M,T)\tau^{1/12}\|\varphi\|_{V_2(Q_\tau)}
\ \ \forall \varphi \in V_2(Q_\tau),
\label{al7}
\end{equation}
\begin{equation}
\|\varphi -\varphi |_{t=0}\|_{C({\overline Q}_\tau)}
\leq c_2(M)\tau^{1/4}\|\varphi\|_{H^{2,1}(Q_\tau)}
\ \ \forall \varphi \in H^{2,1}(Q_\tau)
\label{al8}
\end{equation}
(which follow from the H\"older inequality, the embedding
$V_2(Q_T)\subset L^6(Q_T),$ and the inequality
$\|\phi \|_{C({\overline \Omega})} \leq
c_3(M)\|\phi \|^{1/2}_\Omega \|\phi \|^{1/2}_{H^1(\Omega)}$).
Thus, for $0<\tau\leq \tau_1$ small enough,
\begin{equation}
\|\theta_x\|_{L^4(Q_\tau)}+\|v_x\|_{L^4(Q_\tau)}\leq N_1/2,\ \
(2N)^{-1}\leq \theta\leq 2c_0N\ \ \mbox{in}\ {\overline Q}_\tau.
\label{al9}
\end{equation}

We rewrite the problem for $\eta$ as the integral equation
\begin{equation}
\nu \log \eta=\nu \log \eta^0+I_0(p[\eta,\theta]-p_S)-I^*(v-v^0).
\label{al10}
\end{equation}
For $0<\tau\leq \tau_2$ small enough, this equation has a unique solution
$\eta \in C({\overline Q}_\tau),\, \eta>0,$ and it satisfies the bounds
\begin{equation}
(2N)^{-1}\leq \eta\leq 2c_0N\ \ \mbox{in}\ {\overline Q}_\tau.
\label{al11}
\end{equation}
Moreover, from (\ref{al4}) and (\ref{al10}) it follows that
$\eta_t\in V_2(Q_\tau),$ $\eta\in H^{2,1}(Q_\tau),$ and
\begin{equation}
\|\eta_t \|_{V_2(Q_\tau)}\leq K_7,\
\|\eta_x \|_{L^{2,\infty}(Q_\tau)}\leq K_8,\
\|\eta_{xx} \|_{L^{2,\infty}(Q_\tau)}\leq K_9
\label{al11a}
\end{equation}
(for the last estimate we use conditions (\ref{al1})).
So by applying estimate  (\ref{al7}), for $0<\tau\leq \tau_3$ small enough,
\begin{equation}
\|\eta_x \|_{L^4(Q_\tau)}+\|\eta_t \|_{Q_\tau}\leq N_1/2.
\label{al12}
\end{equation}
In addition, the following estimate holds
\begin{equation}
\sup_{0<\gamma<\tau}\gamma^{-1/2}\|\Delta_{\gamma} \eta_t\|_{Q_{\tau-\gamma}}
\leq K_{10}
\label{al13}
\end{equation}
with $\Delta_{\gamma} \varphi(x,t)=\varphi(x,t+\gamma)-\varphi(x,t).$
This estimate is valid in virtue of the equation
$$
\nu (\log \eta)_t=p[\eta,\theta]-p[\widetilde \eta,\theta]
+\nu \widetilde \rho v_x$$
(where equation (\ref{al3}) is used) and the known estimate
$\sup_{0<\gamma<\tau}\gamma^{-1/2}\|\Delta_{\gamma} \varphi_x\|_{Q_{\tau-\gamma}}
\leq  c_4(M,T)\|\varphi \|_{H^{2,1}(Q_\tau)}$ for all
$\varphi \in H^{2,1}(Q_\tau).$

Thus, for ${\overline \tau}= \min\{\tau_1,\tau_3\},$ the operator $A$ is well defined
and $A(S_{\overline \tau})\subset S_{\overline \tau}$, see  (\ref{al9}), (\ref{al11}), and
(\ref{al12}). Moreover estimates (\ref{al6a}), (\ref{al6b}), (\ref{al11a}), and
(\ref{al13}) imply that the set $A(S_{\overline \tau})$ is precompact in
${\bf B}_{\overline \tau}$.

To prove the continuity of $A,$ take a sequence
$\{ \widetilde z_n\}\subset S_{\overline \tau},$
$\| \widetilde z_n- \widetilde z \|_{{\bf B}_{\overline \tau}}\to 0$ as
$n\to \infty$ and set $z_n=(\eta_n,v_n,\theta_n):=A \widetilde z_n$ and
$z=(\eta,v,\theta):=A \widetilde z.$ By considering problems for
$\theta-\theta_n$ and $v-v_n$, applying the standard parabolic energy estimate
and estimates (\ref{al6a}), (\ref{al6b}), we obtain
$$
\|\theta-\theta_n \|_{V_2(Q_{\overline \tau})}
\leq K_{11}\|{\widetilde z}-{\widetilde z}_n\|_{{\bf B}_{\overline \tau}}\to 0,
$$
$$
\|v-v_n \|_{V_2(Q_{\overline \tau})}
\leq K_{12}(\|{\widetilde z}-{\widetilde z}_n\|_{{\bf B}_{\overline \tau}}
+\|\theta-\theta_n \|_{Q_{\overline \tau}})\to 0.
$$
Considering the difference of equation (\ref{al10}) for $\eta$ and the similar
one for $\eta_n,$ we also obtain
$$
\|\eta-\eta_n \|_{L^{2,\infty}(Q_{\overline \tau})}
\leq K_{13}(\|\theta-\theta_n \|_{Q_{\overline \tau}}
+\|v-v_n \|_{L^{2,\infty}(Q_{\overline \tau})})\to 0.
$$
As the set $A(S_{\overline \tau})$ is precompact, the last three limiting
properties imply that $\|z-z_n\|_{{\bf B}_{\overline \tau}}\to 0.$

Combining all the properties of $S_{\overline \tau}$ and $A$, by the classical Schauder
theorem, we establish that $A$ has a fixed point in $S_{\overline \tau}.$ Evidently this
fixed point serves as a regular weak solution to the original problem
(\ref{e1}), (\ref{e7}), and (\ref{e8}) in $Q_{\overline \tau}.$

Condition (\ref{al1}) can be removed by the standard argument (by smoothing $p_0,p_1$ and $\eta ^0, g$
and passing to the limit).$\ \ \ \Box$

\vskip 0.5cm
 {\bf Remark:}

  In the case $\kappa=\kappa(\eta)$, the existence of $\kappa_{\eta \eta}\in C({\R}^+)$ is not required and the proof can be simplified
  in an essential manner. Namely, the standard parabolic
  $H^{2,1}(Q_T)$-estimates imply $\|\theta\|_{H^{2,1}(Q_T)}\leq K^{(15)}$ in step 3, and
  estimate (\ref{a7}) in step 4 together with the main part of step 6 can be omitted.


\begin{thebibliography}{99}

\bibitem{AZ89}
A.A. Amosov, A.A.Zlotnik,
{\it Global generalized solutions of the equations of the one-dimensional motion of a viscous heat-conducting gas,}
{\em Sov. Math. Doklady} {\bf 38} 1-5 (1989)

\bibitem{AZ92}
A.A. Amosov, A.A. Zlotnik,
{\it Solvability ``in the large" of a system of equations of the
one-dimensional motion of an inhomogeneous viscous heat-conducting gas,}
{\em Math. Notes} {\bf 52} 753-763 (1992)

\bibitem{AZ96}
 A.A. Amosov, A.A. Zlotnik,
{\it Remarks on properties of generalized solutions from $V_2(Q)$ for
one-dimensional parabolic equations,}
{\em MPEI Bulletin} {\bf 3} No.6 15-29 (1996) (In Russian)

\bibitem{AZ97}
 A.A. Amosov, A.A. Zlotnik,
{\it Properties of generalized solutions of one-dimensional linear parabolic
problems with nonsmooth data,}
{\em Differential Equations} {\bf 33} 83-96 (1997)

\bibitem{AB82}
G. Andrews, J.M. Ball,
{\it Asymptotic behaviour and changes of phase in one-dimensional nonlinear viscoelasticity,}
{\em J. Diff. Eqns.} {\bf 44} 306-341 (1982)

\bibitem{AKM}
S.N. Antontsev, A.V. Kazhikhov, V.N. Monakhov,
{\it Boundary value problems in mechanics of nonhomogeneous fluids,}
{\em North Holland} (1990)

\bibitem{D03}
 B. Ducomet,
{\it Global existence for a simplified model of nuclear fluid in one dimension,}
{\em J. Math. Fluid Mech.} {\bf 2} 1-15 (2000)

\bibitem{D01}
 B. Ducomet,
{\it Global existence for a simplified model of nuclear fluid in one dimension: the $T>0$ case,}
{\em Applications of Mathematics} {\bf 47} 45-75 (2002)

\bibitem{D02}
 B. Ducomet,
{\it Simplified models of quantum fluids in nuclear physics,}
{\em Mathematica Bohemica} {\bf 126} 323-336 (2001)

\bibitem{DZ01}
 B. Ducomet, A.A. Zlotnik,
{\it Remark on the stabilization of a viscous barotropic medium with a non-monotone equation of state,}
{\em Applied Math. Letters} {\bf 14} 921-926 (2001)

\bibitem{DZ02}
 B. Ducomet, A.A. Zlotnik,
{\it On the stabilization of a viscous barotropic self-gravitating medium with a non-monotone equation of state,}
{\em Math. Models and Meth. in Appl. Sci.} {\bf 12} 143-153 (2002)

\bibitem{HL98}
L. Hsiao, T. Luo,
{\it Large-time behaviour of solutions to the equations of one-dimensional nonlinear thermoviscoelasticity,}
{\em Quart. Appl. Math.} {\bf 56} 201-219 (1998)

 \bibitem{J}
S. Jiang,
{\it On the asymptotic behavior of the motion of a viscous heat-conducting one-dimensional real gas,}
{\em Math. Z.} {\bf 216} 317-336 (1994)

\bibitem{KAW}
B. Kawohl,
{\it Global existence of large solution to a initial boundary value problems for a viscous heat-conducting one-dimensional real gas,}
{\em J. Diff. Eqns.} {\bf 58} 76-103 (1985)

\bibitem{K90}
 K. Kuttler,
{\it Initial boundary value problems for the displacement in an isothermal, viscous gas,}
{\em Nonlinear Anal. TMA} {\bf 15} 601-623 (1990)

\bibitem{LSU}
O.A. Ladyzhenskaya, V.A. Solonnikov, N.N. Ural'ceva,
{\it Linear and quasilinear equations of parabolic type,}
{\em Amer. Math. Soc., Providence, R.I.} (1968)

\bibitem{Nag}
 T. Nagasawa,
{\it On the outer pressure problem of the one-dimensional polytropic ideal gas,}
{\em Japan J. Appl. Math.} {\bf 15} 53-85 (1988)

\bibitem{P87}
 R. Pego,
{\it Phase transitions in one-dimensional nonlinear viscoelasticity: admissibility and stability,}
{\em Arch. Rat. Mech. and Anal.} {\bf 97} 353-394 (1987)

\bibitem{Q01}
Y. Qin,
{\it Global existence and asymptotic behaviour for a viscous heat-conducting
one-dimensional real gas with fixed and thermally insulated endpoints,}
{\em Nonlinear Anal. TMA} {\bf 44} 413-441 (2001)

\bibitem{RZ97}
R. Racke, S. Zheng,
{\it Global existence and asymptotic behaviour in nonlinear thermoviscoelasticity,}
{\em J. Diff. Eqns.} {\bf 134} 46-67 (1997)

\bibitem{SZZ99}
W. Shen, S. Zheng, P. Zhu,
{\it Global existence and asymptotic behaviour of weak solutions to nonlinear thermoviscoelastic systems with clamped boundary conditions,}
{\em Quart. Appl. Math.} {\bf 57} 93-116 (1999)

\bibitem{SZ01}
I. Stra\v skraba, A. Zlotnik,
{\it On a decay rate for 1d viscous compressible barotropic fluid equations,}
{\em To appear in: J. Evol. Equat.} {\bf 1} (2001)

\bibitem{Z92}
 A.A. Zlotnik,
{\it On equations for one-dimensional motion of a viscous barotropic gas in the presence of a body force,}
{\em Siberian Math. J.} {\bf 33} 798-815 (1992)

\bibitem{Z00}
 A.A. Zlotnik,
{\it Uniform estimates and the stabilization of symmetric solutions to one system of quasilinear equations,}
{\em Differential Equations} {\bf 36} 701-716 (2000)

\bibitem{ZB94}
 A.A. Zlotnik, N.Z. Bao,
{\it Properties and asymptotic behaviour of solutions
of some problems of one-dimensional motion of a viscous barotropic gas,}
{\em Math. Notes} {\bf 55} 471-482 (1994)


 \end{thebibliography}
    \end{document}